\documentclass[twocolumn,american,5p, twocolumn]{revtex4-2}
\usepackage[T1]{fontenc}
\usepackage[utf8]{inputenc}
\usepackage{xcolor}
\usepackage{babel}
\usepackage{array}
\usepackage{float}
\usepackage{booktabs}
\usepackage{amsmath}
\usepackage{amssymb}
\usepackage{graphicx}
\PassOptionsToPackage{normalem}{ulem}
\usepackage{ulem}
\usepackage[pdfusetitle,
 bookmarks=true,bookmarksnumbered=false,bookmarksopen=false,
 breaklinks=false,pdfborder={0 0 1},backref=false,colorlinks=true]
 {hyperref}

\makeatletter

\providecommand{\tabularnewline}{\\}
\providecolor{lyxadded}{rgb}{0,0,1}
\providecolor{lyxdeleted}{rgb}{1,0,0}
\DeclareRobustCommand{\mklyxadded}[1]{\textcolor{lyxadded}\bgroup#1\egroup}
\DeclareRobustCommand{\mklyxdeleted}[1]{\textcolor{lyxdeleted}\bgroup\mklyxsout{#1}\egroup}
\DeclareRobustCommand{\mklyxsout}[1]{\ifx\\#1\else\sout{#1}\fi}

\usepackage{amssymb}

\makeatother

\begin{document}

\title{Phase Transition From Turbulence To Zonal Flows In The Hasegawa-Wakatani
System}
\author{P. L. Guillon$^{1,2}$ and Ö. D. Gürcan$^{1}$}
\affiliation{$^{1}$Laboratoire de Physique des Plasmas, CNRS, Ecole Polytechnique,
Sorbonne Université, Université Paris-Saclay, Observatoire de Paris,
F-91120 Palaiseau, France}
\affiliation{$^{2}$École des Ponts, 77455 Marne-la-Vallée Marne-la-Vallée cedex 2, France}

\begin{abstract}
The transition between two-dimensional hydrodynamic turbulence and
quasi-one-dimensional zonostrophic turbulence is examined in the modified
Hasegawa-Wakatani system, which is considered as a minimal model of
$\beta$-plane-like drift-wave turbulence with an intrinsic instability.
Extensive parameter scans were performed across a wide range of values
for the adiabaticity parameter $C$ describing the strength of coupling
between the two equations. A sharp transition from 2D isotropic turbulence
to a quasi-1D system, dominated by zonal flows, is observed using
the fraction of the kinetic energy of the zonal modes as the order
parameter, at $C\approx0.1$. It is shown that this transition exhibits
a hysteresis loop around the transition point, where the adiabaticity
parameter plays the role of the control parameter of its non-linear
self-organisation. It was also observed that the radial particle flux
scales with the adiabaticity parameter following two different power
law dependencies in the two regimes. A simple quasi-linear saturation
rule which accounts for the presence of zonal flows is proposed, and
is shown to agree very well with the observed nonlinear fluxes. Motivated
by the phenomenon of quasi-one dimensionalisation of the system at
high $C$, a number of reduction schemes based on a limited number
of modes were investigated and the results were compared to direct
numerical simulations. In particular, it was observed that a minimal
reduced model consisting of $2$ poloidal and $2$ radial modes was
able to replicate the phase transition behaviour, while any further
reduction failed to capture it.
\end{abstract}
\maketitle

\section{Introduction}

The two dimensional Hasegawa-Wakatani system, which describes non-linear
dynamics of dissipative drift-wave instability \citep{hasegawa:83}, appears
as a minimal model of drift wave turbulence in tokamak plasmas. Despite
its simplicity, this model already displays complex nonlinear dynamics,
including various feedback mechanisms such as the interplay between
turbulence and zonal flows \citep{gurcan:15}. Studies of this system
are primarily motivated by the need for a better understanding of
the mechanisms of self-regulation of drift wave turbulence (and other
instabilities) through the formation of zonal flows, and the impact
of this co-evolution on the particle flux. While the Hasegawa-Wakatani
model is not directly applicable to tokamaks, the nonlinear feedback
mechanisms, that eventually determine the radial fluxes in this system,
are probably also active in more realistic models, where they may
have a direct impact on the efficiency of future fusion reactors \citep{diamond:2005,itoh:2006,miki:12}.
In particular, the development of models that are further reduced,
but capable of capturing the complexity of the full two dimensional
Hasegawa-Wakatani system, can be very useful in order to identify
the reduction schemes that allow a proper description of these key
mechanisms, which can then be applied to improve realistic turbulent-transport
models of magnetic confinement devices \citep{baschetti:2021}.

In this article, we explore the dependency of zonal flow formation
on the linear parameters of the system, mainly the ratio between the
adiabaticity constant $C$, which couples the two equations, and the
background density gradient $\kappa$, which provides the free energy
source. Earlier studies of the so-called modified Hasegawa-Wakatani
equations, which is actually the proper 2D version of the three dimensional
Hasegawa-Wakatani system, showed that it responds in qualitatively
different ways depending on the values of these two parameters. Low
values of $C/\kappa$ correspond to an eddy dominated state, close
to 2D isotropic turbulence, whereas $C/\kappa\gtrsim1$ leads to the
formation of zonal flows, which are stationary, quasi-periodic large-scale
radial structures \footnote{Usually, studies of the dependency on $C$ and $\kappa$ are carried
out independently, and the name of the regimes correspond mostly to
the limits of $C\rightarrow0$ (\emph{hydrodynamic}) and $C\rightarrow +\infty$
(\emph{adiabatic}) due to the coupling nature of the adiabaticity
parameter. However, as we explain in Section \ref{subsec:Linear-solutions},
the reasoning prevails also using $C\rightarrow C/\kappa$ for the
sake of generality.}. Non-linear formation of such an effectively 1D pattern can be seen
as an example of self-organisation, through a mechanism of predator-prey
dynamics with zonal flows feeding on turbulence \citep{malkov:01,kobayashi:15}.
The transition point between these two limiting regimes is believed
to be of order $C/\kappa\sim0.1-1$ \citep{numata:2007} (with a normalised
density gradient $\kappa=1$). Recently, it was pointed out in Ref.
\onlinecite{grander:2024} that a qualitative \emph{hysteresis} behaviour
is observed in the collapse and re-generation of zonal flows, as $C$
is varied across the transition point. A similar behaviour was also
observed in turbulence driven by trapped particles in gyrokinetic
simulations \citep{gravier:2017}. However in both works, the parameter
sweep performed to investigate the hysteresis was done relatively
quickly (of the order of a few inverse linear growth rates), and the
behaviour that is observed is thus indistinguishable from the ``memory''
(or inertia) of the turbulent system, which naturally requires some
time to adapt to the parameter change. Thus, one still needs to establish the dynamics of the transition and make the connection to the physics of phase transitions, as we will begin to address in the following.

\begin{figure}[htbp]
\begin{centering}
\includegraphics[width=1\columnwidth]{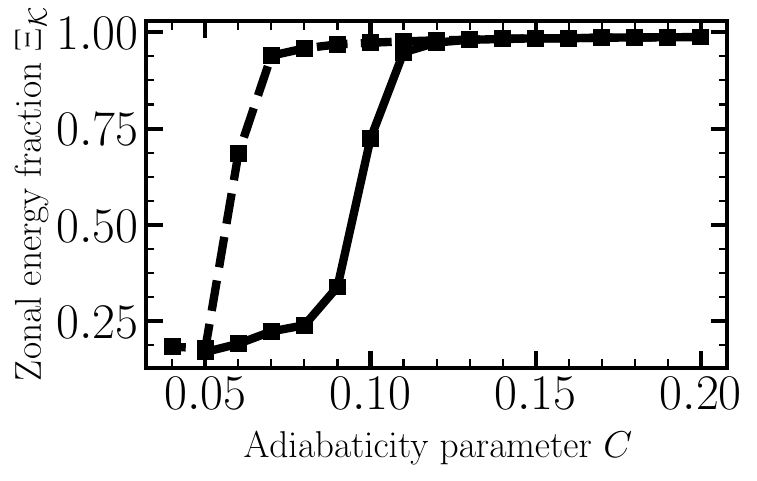}
\par\end{centering}
\caption{Hysteresis exhibited by the zonal energy fraction $\Xi_{K}$ (ratio
of the zonal to total kinetic energy) in a simulation where the adiabaticity
parameter $C$ is increased (solid black) and then decreased (dashed
black) across the transition point. \label{fig:HYST-INTRO}}
\end{figure}

In this work we study the transition from an eddy dominated, two dimensional
turbulence state to the zonal flow dominated state (or vice versa)
using the language of phase transitions, with a clearly identified
control parameter $C/\kappa$ and a proposed effective order parameter,
which is the ratio of the zonal to total kinetic energy (hereafter
referred to as ``zonal energy fraction'' and denoted $\Xi_{K}$).
Using these variables, and performing an extensive parameter scan
of the adiabaticity parameter $C$ in the range $[10^{-4},20]$, we
recover the transition observed in Ref. \onlinecite{numata:2007}, with
an increased resolution and number of data points. We also demonstrate,
for the first time, the actual hysteresis loop for the zonal energy
fraction, as can be seen in Figure \ref{fig:HYST-INTRO}, when the
adiabaticity parameter is slowly increased and then decreased across
the transition point, waiting long enough at each increment to let
the system adapt to the new value of $C$ and ``forget'' its previous
state (\emph{i.e.} performing an ``adiabatic'' transformation in
the thermodynamic sense). Looking at the hysteresis as a feature of
a phase transition between a ``hot'' disordered state (isotropic
2D turbulence) and a ``colder'' organised state (zonal flows), with
$(C/\kappa)^{-1}$ serving as a proxy for heating, this observation
suggests that, once zonal flows are established, their collapse requires
some energy absorption akin to \emph{latent heat}, suggesting a first
order transition. Indeed, the existence of the upper branch for the
backtransition may be akin to crystal melting, also suggested in the
context of staircase vortex melting \citep{ramirez:2024}.

In this perspective, the formation of strong zonal structures from
turbulence can also be seen as a transition from 2D isotropic turbulence
to a quasi-1D state dominated by sheared flows, since the system becomes
almost one dimensional, where the variation is only in the radial
direction. Analogous transitions have been observed in rotating magnetohydrodynamic
turbulence for the solar tachocline \citep{tobias:07}, or 3D astrophysical
flows \citep{benavides:2022}. Note that transitions from 3D to quasi-2D
flow have also been observed in thin layer turbulence \citep{benavides:2017,alexakis:2018},
for which the Navier-Stokes equation is evolved in a 3D box where
the height $H$ is very small compared to the two other dimensions.
It has been argued however, that such systems exhibit second-order
phase transitions of the energy contained in the large quasi-2D scales
versus energy contained in the small 3D scales while varying the ratio
between the length scale $l_{f}$ of the forcing and the thickness
$H$ of the layer. This is believed to be the result of a competition
between the 2D inverse cascade and the 3D direct cascade. The former
is predominant when energy is injected at scales larger than the thickness
of the layer and forms large, quasi-2D structures, while for $l_{f}\ll H$
the energy cascades towards smaller scales. The formation of large-scale
2D condensates in those systems \citep{kan:2019,kan:2019a}, or in
rotating turbulent flows \citep{seshasayanan:2018}, can be subcritical
and exhibit a hysteresis behaviour. The transition between 2D condensates
and 1D zonal jets in 2D Navier-Stokes has recently been studied in
Ref. \onlinecite{xu:2024}. Similar phase transitions have also been observed and characterised
in experimentents involving a highly turbulent closed flow \citep{cortet:2011}.

In the case of the Hasegawa-Wakatani system, we are faced with a transition
that depends on the energy injection mechanism, since $C$ and $\kappa$
control the linear instability which drives the turbulence. The fact
that these linear parameters constrain the formation of non-linear
structures suggests that the transfer of energy between the linear
instability, turbulence and zonal flows is governed by a competition
between linear and non-linear mechanisms. Such dependency can be partially
understood using the zonostrophy parameter \citep{gurcan:2024,scott:2012},
which can be defined for the Hasegawa-Wakatani system as: $R_{\beta}\equiv k_{c}/k_{peak}$.
Notice that here, $k_{c}=C/\kappa$ is the critical wave-number, analogous
to the transition scale for $\beta$-plane turbulence, which separates
the adiabatic (or highly zonostrophic) behaviour for large scales
from the hydrodynamic behaviour at small scales, and $k_{peak}$ is
the wave-number at which the kinetic energy spectrum is peaked. When
$R_{\beta}>1$, the large-scale zonal modes are dominant. On the contrary,
when $R_{\beta}\ll1$, $k_{peak}$ is located in the bulk of the turbulent
spectrum and the system is in an eddy dominated state. Considering
that $k_{peak}\sim0.5-1$ (corresponding either to the most unstable
mode or the dominant zonal mode), we see that $R_{\beta}$ scales
like $C/\kappa$, which explains why the latter controls the fate
of the zonal flows. However, such analysis relies on dimensional arguments
and cannot explain neither the numerical value of the critical control
parameter $C/\kappa\approx0.1$, at which the system transitions,
nor the nature of the transition, whether it is smooth or abrupt,
and in particular the phenomenon of hysteresis that is observed around
the transition point.

Another way to look at this transition as an extension of the transition
from three to two dimensions and then to quasi-1D state, can be seen
in terms of the number of positive definite conserved quantities.
It is well known that in two dimensions, the existence of two positive
definite conserved quantities (energy and enstrophy) forces the system
towards inverse cascade through something like the Fjørtoft argument
\citep{fjortoft:1953}. However the transition from forward to inverse
cascade in three dimensions (hence what we characterise as quasi-2D
turbulence) actually occurs before enstrophy is exactly conserved
in the three dimensional flow. Similarly, we can argue that in the
adiabatic limit of the Hasegawa-Wakatani system, corresponding to
the wave-turbulence regime in the Charney-Hasegawa-Mima model, a third
conservation law emerges \citep{balk:1991,balk:1991b}: the pseudo-momentum
(also called zonostrophy \citep{nazarenko:2009,connaughton:2015}).
The presence of this third conserved quantity forces the wave-wave
interactions to transfer their energy towards zonal modes. The existence
of the conservation law in the asymptotic limit plays a role, even
before this quantity is fully conserved, much like the transition
to two dimensional turbulence. In this sense, the phase transition
that is discussed here can also be seen as a transition from strong
two dimensional turbulence dominated by eddies, to ``weak'' wave
turbulence, dominated by wave-wave interactions eventually giving
their energy to zonal flows \citep{connaughton:2015,newell:2011,galtier:2022}.
While the Hasegawa-Wakatani system is not a wave-turbulence problem
\emph{per se}, the transfer of energy to finite $k_{x}$, $k_{y}\approx0$
modes with varying $C$ can be studied in the same spirit \citep{pushkarev:13}.

As a consequence of the transition, it was observed that the particle
flux displayed a power law behaviour in $C$, and while in the eddy
dominated phase, this scaling had the form $\Gamma\propto C^{-0.35}$,
in the zonal flow dominated phase, it fell off much sharper with $\Gamma\propto C^{-2}$,
with the same transition point, around $C\approx0.1$, following exactly
the same asymptotic scalings suggested in Ref. \onlinecite{hu:97}, which
were obtained for the standard (\textit{i.e.} non-modified) Hasegawa-Wakatani
system, with a sharper transition in our case. Note that as we approach
the adiabatic limit, the flux becomes very small and hard to trace
because of meandering and merging of zonal flows. Comparing the nonlinear
flux to a simple quasi-linear estimation using the saturation rule,
multiplied by the turbulent energy fraction $1-\Xi_{K}$, where $\Xi_{K}$
is the fraction of kinetic energy in the zonal modes, in order to consider the fact
that only non-zonal modes contribute to particle flux, we find a very
good agreement. In particular, we observe that the analytical expression
for the linear growth rate $\gamma_{k}$ (in fact $\left.\gamma_{k}/k^{2}\right|_{\text{max}}$
but the two are similar in the limit) for $C\ll1$, when substituted
into the saturated flux scales as $C^{-1/3}$, which is close to the
measured exponent. We think that this provides an interesting way
to incorporate the effects of zonal flows in quasi-linear modelling
also in gyrokinetic models.

Last, we were able to observe the transition in a truncated reduced
model composed of only 12 modes, involving two poloidal modes corresponding
to the most unstable mode $k_{y0}$ and its first subharmonic (\textit{i.e.}
$k_{y0}/2$) together with two zonal modes and the necessary sidebands.
We think that this is the minimal reduced model that is able to reproduce
the transition, since it contains a triadic interaction between turbulent
modes and the zonal modes, but at the same time some turbulent-turbulent
interactions reproducing the inverse energy cascade. This is confirmed
by the fact that a further reduced system, with only 1 poloidal mode,
or a system with the same number of modes but with the first harmonic
(\textit{i.e.} $2k_{y0}$) instead of the subharmonic, was not able able to
reproduce the transition, because the ``inverse cascade'' mechanism
between turbulent modes is removed in these reductions.

The rest of the article is organised as follows. First, we give a
description of the Hasegawa-Wakatani model in Section \ref{sec:The-Hasegawa-Wakatani-equations},
where we detail the linear properties and particularly the energy
injection mechanism. We also give some illustration of the non-linear
behaviours of the system, and define several quantities that are useful
to study the transition, mainly the fraction of zonal kinetic energy
and zonal enstrophy of the system, as well as the radial particle
flux. In Section \ref{sec:C-dep}, we present the results of the parameter
scan, with $C\in[10^{-4},20]$. The zonal energy fraction displays
a transition from a fully turbulent system to strong zonal flows at
$C\approx0.1$ (with $\kappa=1$), while the enstrophy fraction gradually
increases at the transition point. We also evidence two different
power law dependencies of the particle flux to the adiabaticity parameter,
depending on the regime the system is in, and compare the flux to
a formulation using a simple saturation rule accounting for the effects
of zonal flows yielding good qualitative and quantitative agreements.
We then investigate the behaviour of the system around the transition
point and evidence the hysteresis that both energy and enstrophy zonal
fractions exhibit when increasing and then decreasing the adiabaticity
parameter during a single simulation run. Finally, in Section \ref{sec:Reductions},
we try to reproduce our observations using reduced models involving
only a few Fourier modes. We find that, in order to reproduce the
phase transition, a minimal reduction should involve (at least) 2
poloidal modes corresponding to scales larger than the energy injection
scale.

\section{The Hasegawa-Wakatani equations\label{sec:The-Hasegawa-Wakatani-equations}\label{sec:The-Hasegawa-Wakatani-equations}}

\subsection{Description of the system}

The Hasegawa-Wakatani model describes turbulence generated by the
dissipative drift-wave instability, driven by a ``background'' density
gradient in the radial direction, where the fluctuations are in a
plane orthogonal to a uniform and constant magnetic field $B$. In
the following, the spatial variables $x$ and $y$ are normalised
to $\rho_{s}$ the sound Larmor radius and time $t$ is normalised
to $\Omega_{i}$ the cyclotron frequency. The $x$ direction corresponds
to the radial direction in a tokamak, while $y$ corresponds to the
poloidal direction. The modified - but really, the only correct -
2D version of the Hasegawa-Wakatani equations can be written as \citep{hasegawa:83,gurcan:2022}
\begin{align}
\partial_{t}\nabla{{}^2}\phi+[\phi,\nabla{{}^2}\phi] & =C(\widetilde{\phi}-\widetilde{n})+\nu\nabla^{4}\widetilde{\phi}\label{eq:hw1}\\
\partial_{t}n+[\phi,n]+\kappa\partial_{y}\phi & =C(\widetilde{\phi}-\widetilde{n})+D\nabla^{2}\widetilde{n}\label{eq:hw2}
\end{align}
Here $\phi$ is the electric potential normalised to $T/e$, where
$T$ is the electron temperature and $e$ is the unit charge, $\nabla^{2}\phi$
is the vorticity and $n$ is the density perturbation, normalised
to a background reference density $n_{0}$. The Poisson bracket is
defined as $[\phi,A]\equiv(\boldsymbol{\hat{z}}\times\nabla\phi)\cdot\nabla A$
and describes the advection of the quantity $A$ by the $E\times B$
drift velocity, defined in these dimensionless variables as: $\mathbf{v}_{E\times B}\equiv\boldsymbol{\hat{z}}\times\nabla\phi$,
with $\boldsymbol{\hat{z}}$ the unit vector parallel to the magnetic
field. The background density gradient is given by $\kappa\equiv-\frac{1}{n_{0}}\frac{dn_{0}}{dx},$
where $n_{0}\left(x\right)$ is a background density profile in the
radial direction, assumed to be linearly decreasing with $x$. The
adiabaticity parameter $C$, which allows the two equations to be
coupled, can be written as $C=\frac{k_{||}^{2}\rho_{s}B}{en_{0}\eta_{e,||}}$,
with $k_{||}$ the parallel wave-number of fluctuations (along the
magnetic field lines) and $\eta_{e,||}$ the parallel resistivity.
The main interest in studying the Hasegawa-Wakatani system is its
ability to describe the self-organisation of instability-driven turbulence
into large-scale, poloidal flows alternating along the radial direction,
called \emph{zonal flows}. In wave-number space, they are associated
with $k_{y}=0$ modes. We can decompose each perturbation $A$ into
its zonal and non-zonal parts: $A=\overline{A}+\widetilde{A}$, where
$\overline{A}\equiv\langle A\rangle_{y}=\frac{1}{L_{y}}\int_{0}^{L_{y}}Ady$\emph{
}is averaged along the $y$ direction, and $\langle\widetilde{A}\rangle_{y}=0$.
The coupling term $C(\widetilde{\phi}-\widetilde{n})$ applies only
on the non-zonal component of the fluctuations since zonal flows are
axisymmetric structures and hence have $k_{\parallel}=0$.

When $C\rightarrow0$, the two equations decouple and the vorticity
equation (\ref{eq:hw1}) becomes the the 2D incompressible Navier-Stokes
equation ($\phi$ plays the role of the stream function in an incompressible
flow), while the density equation (\ref{eq:hw2}) becomes a passive
scalar equation with a background gradient. This limit where the two
equations are decoupled is called the \emph{hydrodynamic regime}.
Note that another possible solution in this limit, with some non-zero
constant electrostatic potential, is the so-called non-normal mode
and a secularly growing density perturbation \citep{camargo:1998}.
On the other hand, when $C\rightarrow +\infty$, the\emph{ }coupling
term $C(\widetilde{\phi}-\widetilde{n})$ dominates and forces $\widetilde{\phi}=\widetilde{n}$
at the leading order. This limit is called the \emph{adiabatic} \emph{regime},
because $C\rightarrow +\infty$ means that the electron parallel resistivity
$\eta_{e,||}$ goes to 0 causing the perturbations of the electrostatic
potential to be in phase with the fluctuations of density. In this
limit, the system no longer presents linear instability, but only
drift-waves propagating in the $+y$ direction, and one has to force
it in order to generate turbulence. Since $\widetilde{\phi}=\widetilde{n}$,
the equations can be subtracted and condensed into the well known
form of the Charney-Hasegawa-Mima equation, which describes both drift-waves
in magnetised plasmas or Rossby waves in rotating planetary atmospheres
\citep{gurcan:15,hasegawa:1978}.

Even though we are interested in the inviscid limit of the Hasegawa-Wakatani
system, and in particular the dynamics of the zonal flows, generated
by an initial linear instability, we introduce small-scale dissipation
terms on non-zonal fluctuations through the viscosity coefficients
$\nu$ and $D$, acting respectively on vorticity $\nabla^{2}\widetilde{\phi}$
and density $\widetilde{n}$ non-zonal fluctuations. These diffusion
terms are mainly used for numerical purposes, in order to balance
the exponential energy injection by the linear instability, and avoid
the accumulation of energy which reaches the smallest scales by the
direct enstrophy cascade. Note that no dissipation or friction (small
or large scale) is applied on zonal modes. As a result the energy
that is injected by the linear instability can either be transferred
from non-zonal fluctuations to large-scale zonal flows ($k_{y}=0$,
small $k_{x}$ modes), or dissipated by small-scale non-zonal fluctuations
through the terms described previously. Indeed, no accumulation of
energy is observed in the numerical simulations for high $k_{x}$
(small radial scales) zonal modes. Furthermore, since the energy is
mostly concentrated in the large scale zonal flows, it is not interesting
to have small scale dissipation for zonal modes (and the actual small-scale
dissipation mechanism is different from that of turbulent modes).
Obviously, dissipation for both turbulence and zonal flows in tokamak
plasmas is more complex and requires a kinetic approach, particularly
in order to properly take into account the Landau damping on turbulence
and neoclassical effects on zonal flows \citep{hinton:99,gurcan:15},
which is beyond the scope of the present work.

For the sake of generality, we can also introduce a simple form of
large-scale friction on zonal vorticity $\nabla^{2}\overline{\phi}$
and density $\overline{n}$ using the friction coefficients $\nu_{ZF}$
and $D_{ZF}$. In systems exhibiting a 2D inverse cascade of energy,
large-scale friction allow the dissipation of energy reaching the
largest scale, which would otherwise be accumulated at this scale.
Such dissipation mechanism is used for example in $\beta$-plane turbulence
and has been provided as an explanation to the observed size of the
long-lived zonal jets \citep{vallis:1993}. In our simulations, however,
we observe the formation of long-lived zonal flows smaller than the
box size (\emph{i.e. }the largest scale available) \emph{without any
large scale friction term}. We believe that this is due to the ability
of the system to turn-off the linear instability drive, hence the
energy injection. Therefore, we do not consider large-scale friction
terms in the following, which would otherwise constitute an additional
linear parameter, and we focus on the minimal feasible model.

\subsection{Linear properties \label{subsec:Linear-solutions}}

We describe briefly the linear properties of the system. First, the
2D spatial Fourier transform of the Hasegawa-Wakatani equations for
the non-zonal fluctuations, can be written as
\begin{align}
\partial_{t}\phi_{k} & =-\frac{C}{k^{2}}(\phi_{k}-n_{k})-\nu k^{2}\phi_{k}\label{eq:hw1-linf}\\
\partial_{t}n_{k}+i\kappa k_{y}\phi_{k} & =C(\phi_{k}-n_{k})-Dk^{2}n_{k}\text{ ,}\label{eq:hw2-linf}
\end{align}
where $\phi_{k}$ and $n_{k}$ are the Fourier coefficients of the
electric potential and density fluctuations associated to the wave-number $k$.
Note that from now on, $k=(k_{x},k_{y})$ will mainly denote non-zonal
flucutations wave-numbers, whereas $q=(q_{x},0)$ will be preferred
for zonal modes. In Fourier space, amplitudes of zonal modes will
be written with an overline, as in real space. Linearisation and Fourier
transform of the zonal equations yields $\partial_{t}\overline{\phi}_{q}=\partial_{t}\overline{n}_{q}=0$,
which emphasises their non-linear nature. It can be shown \citep{gurcan:2022}
that the system admits 2 eigenvalues, which we denote as $\omega_{k}^{\pm}(C,\kappa,\nu,D)=\omega_{k,r}^{\pm}+i\gamma_{k}^{\pm}$
(with $\omega_{k,r}^{\pm}$ the real frequency of $\omega_{k}^{\pm}$
and $\gamma_{k}^{\pm}$ the growth rate), and which are functions
of the system parameters $C,\kappa,\nu$ and $D$. The expression of the eigenvalues are given in Appendix \ref{sec:app-eigenvalues}. The $"+"$ mode
is associated with a positive growth rate and is thus unstable, while
the $"-"$ mode is damped. Naturally the linear studies below focus
on the unstable mode, since the instability growth rate $\gamma_{k}^{+}$
correponds to the energy injection mechanism of the system. It is
maximum for $k_{x}=0$ modes and for a finite value $k_{y0}$, which
corresponds to the scale at which the energy is initially injected.
In the following, we will use $k_{y0}^{-1}$ as a proxy for the injection
scale.

\begin{figure}[htbp]
\begin{centering}
\includegraphics[width=1\columnwidth]{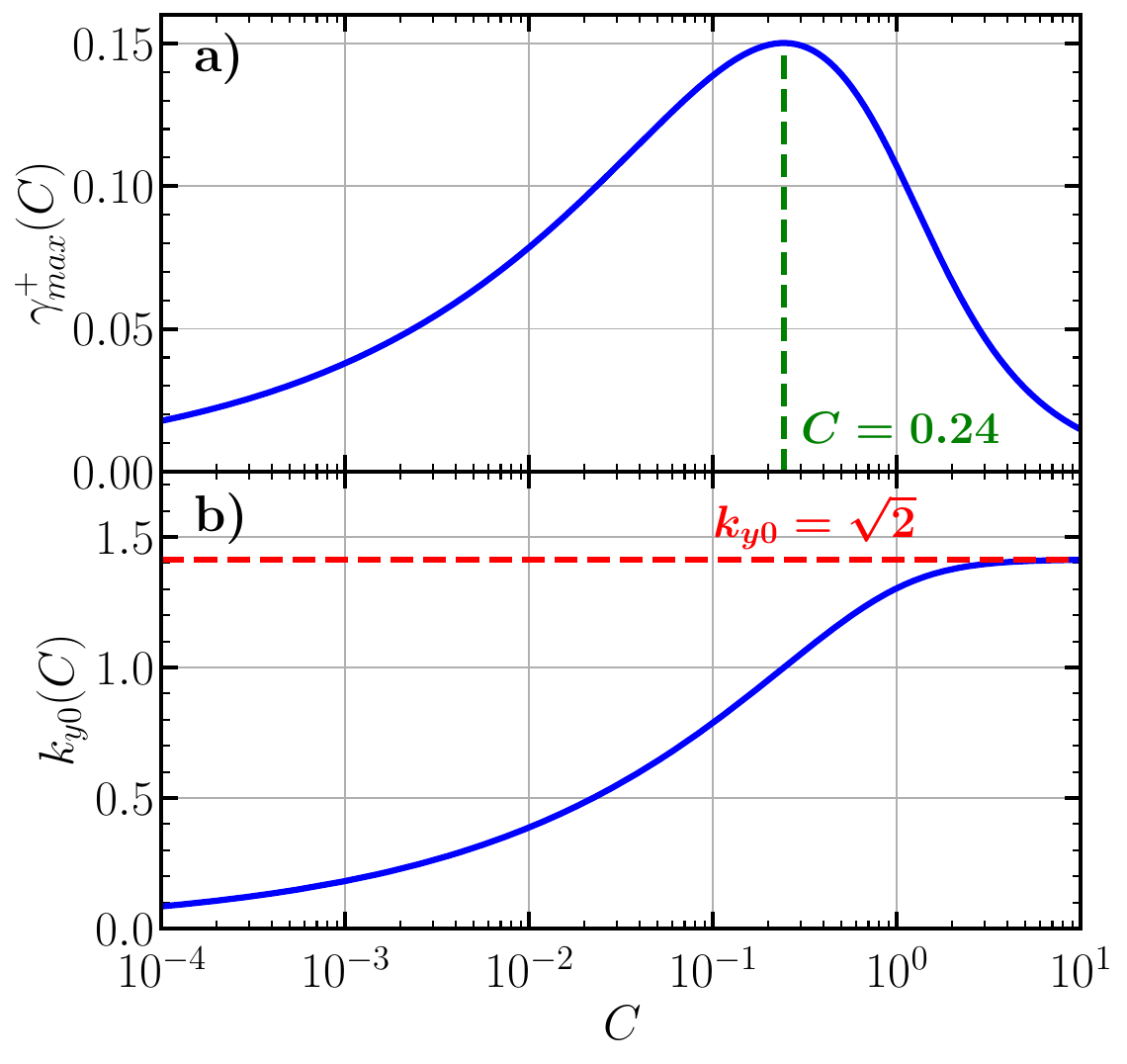}
\par\end{centering}
\caption{Maximum linear growth rate $\gamma_{max}^{+}$ (a) and corresponding
wave-number $k_{y0}$ (b) as functions of the coupling parameter $C$
(semilog scaled), with $(\kappa,\nu,D)=(1,0,0)$. The green dashed
line corresponds to $C\approx0.24$ at which $\gamma_{max}^{+}$ is
maximum. The red dashed line correspond to the limit value of $k_{y0}=\sqrt{2}$
when $C\rightarrow +\infty$. \label{fig:Maximum-linear-growth}}
\end{figure}

The maximum growth rate $\gamma_{max}^{+}$ and the corresponding
wave-number $k_{y0}$ are computed numerically using the expressions
of the eigenvalues given in Appendix \ref{sec:app-eigenvalues}, and
are shown as functions of $C$ in the top and bottom plots of Figure
\ref{fig:Maximum-linear-growth} respectively. It can be seen that
in both hydrodynamic ($C\rightarrow0$) and adiabatic limits ($C\rightarrow +\infty$),
$\gamma_{max}^{+}\rightarrow0$, which means that the instability
disappears in those limits. The maximum growth rate reaches its highest
value for $C\approx0.24$ (green dashed line), taking $\kappa=1$.
Increasing $C$ after this maximum results in a rapid fall of $\gamma_{max}^{+}$,
which corresponds to a decrease of the energy injection rate. From
the point of view of the transition between a hot disordered state
to a cold organised state, this is similar to reducing heating. On
the bottom plot, the injection wave-number $k_{y0}$ goes from very
small values for $C\rightarrow0$ to $\sqrt{2}$ for $C\rightarrow +\infty$.
In comparison, when they form, zonal flows have typically a wave-number
about $1/3$ to $1/2$ of the injection wave-number (this is further
discussed in Section \ref{sec:C-dep}).

The effect of viscosity on the growth rate, especially for small values
of viscosity, is such that it hardly changes its value or the wave-number
at which it is maximum. Since the diffusion operators scale like $k^{2}$
in Fourier space and viscosity coefficient are of order $\nu\sim10^{-3}-10^{-4}$
in our simulations, only the high wave-numbers (small scales) are
affected, for which the growth rate can become negative with finite
viscosity, instead of going to zero as in the inviscid case. Since
we are interested in the large-scale inviscid behaviour of the system,
diffusion is introduced mainly for numerical purposes in order to
balance the exponential energy injection due to linear instability.

Finally, we discuss the effect of $\kappa$ on the linear properties
of the system. As it corresponds to the slope of the background density
gradient, it represents the ``free energy'' source that drives the
linear instability. Hence, the growth rate increases with $\kappa$.
Indeed, a higher $\kappa$ corresponds to a steeper background density
profile, which makes the instability develop faster in order to relax
the system further away from the equilibrium. Dividing both Hasegawa-Wakatani
equations (\ref{eq:hw1})-(\ref{eq:hw2}) by $\kappa{{}^2}$ and letting
\[
\frac{\phi}{\kappa}\rightarrow\phi,\,\frac{n}{\kappa}\rightarrow n\,,\kappa t\rightarrow t\,,
\]
 we get the same equations with $\kappa$ replaced by $1$ and with
a new coupling term $C/\kappa$, which remains the only relevant parameter
to study. Note that diffusion coefficients have also been rescaled
by $\kappa$. Such normalisation is effectively used by several groups
working on the Hasegawa-Wakatani model \citep{camargo:95,hu:1997,kim:2019}.
Therefore in the following, we can study the behaviour of the system
as a function of $C$, by keeping $\kappa=1$ constant without any
loss of generality, and we can argue that it corresponds in fact to
a $C/\kappa$ scaling of this properly normalised system.

\subsection{Non-linear behaviour: eddy and zonal flow dominated states \label{subsec:Non-linear-behaviour:-hydrodynam}}

We now discuss the non-linear dynamics of the Hasegawa-Wakatani model,
and particularly the difference between the\emph{ }eddy dominated state,
close to 2D isotropic turbulence, and the zonal flow dominated state,
where the system can be considered as quasi-1D. We also define some
relevant quantities to study and characterise the system in these
different regimes: fractions of zonal energy and enstrophy, and the
radial particle flux, the latter being of most interest for the study
of turbulent transport in fusion plasmas. To illustrate the discussion,
we show results from high resolution simulations using a pseudo-spectral
Hasegawa-Wakatani solver with a padded resolution of $4096\times4096$.
Some details of the simulations are given in Table \ref{tab:4096_params}.
A complete description of the simulation set-up is given at the begining of
Section \ref{sec:C-dep}, where we give the result of the extensive
adiabaticity parameter scan that we performed.

\begin{table}[h]
\begin{centering}
\begin{ruledtabular}
\begin{tabular}{ccccc}
$C$ & 0.01 & 0.1 & 1 & 10\tabularnewline
\midrule
\addlinespace[0.1cm]
$L_{x},L_{y}$ & 162.3 & 79.7 & 48.2 & 44.5\tabularnewline\addlinespace[0.1cm]
$\nu$ & $2.6\times10^{-3}$ & $1.1\times10^{-3}$ & $3.1\times10^{-4}$ & $3.7\times10^{-5}$\tabularnewline
\end{tabular}
\end{ruledtabular}
\par\end{centering}
\caption{Parameters of the high resolution simulations with a padded resolution
of $4096\times4096$ for some values of $C$. For all results from
$4096^{2}$ simulations presented in this work, the background density
gradient is fixed at $\kappa=1$, the size of the box is $L_{x}=L_{y}=2\pi\times10/k_{y0}^ {}$,
the dissipation is $\nu=0.005\times\gamma_{max}/k_{y0}^{2}$, and
simulations are ran up to $t=50\times\gamma_{max}^{-1}$.\label{tab:4096_params}}
\end{table}

\subsubsection{Behaviour of the system in the two regimes}

\begin{figure*}[htbp]
\begin{centering}
\includegraphics[width=2\columnwidth]{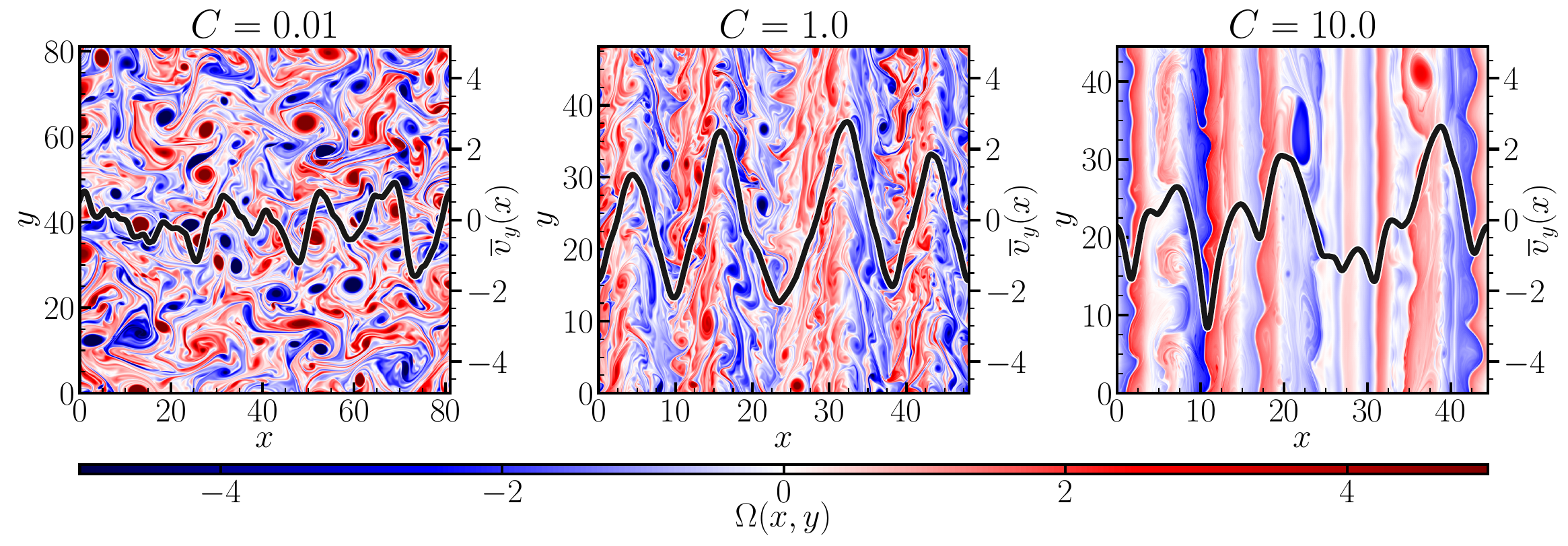}
\par\end{centering}
\caption{Snapshots of vorticity $\Omega(x,y)$ for $C=0.01$ (left), $C=1$
(middle) and $C=10$ (right), obtained from simulations with a padded
resolution of $4096\times4096$. The zonal velocity $\overline{v}_{y}(x)$
as a function of the radial coordinate $x$ is also shown (black line).\label{fig:Snapshots-of-vorticty}}
\end{figure*}

First, we illustrate the different behaviours of the system, as a
function of the adiabaticity parameter $C$. In Figure \ref{fig:Snapshots-of-vorticty},
we show three snapshots of the vorticity field $\Omega(x,y)\equiv\nabla^{2}\phi$
for $C=0.01$ (left), $C=1$ (middle) and $C=10$ (right), taken many
linear growth times after the system has reached non-linear saturation.
We also plot the zonal velocity (black line), defined as $\overline{v}_{y}(x)\equiv\partial_{x}\overline{\phi}$,
which corresponds to the $y$ component of the flow velocity averaged
along $y$. On the left, for $C=0.01$, we see a chaotic system displaying
eddies of different sizes without any clear pattern, which is characteristic
of 2D isotropic turbulence. This is also witnessed by the noisy zonal
velocity profile, which exhibits random small scales features and
fluctuates strongly in time, which is basically the projection of
a turbulent 2D velocity field on the $y$ direction.

In contrast, in the middle and right plots, we see large-scale radial
structures, which in turn correspond to smoother, quasi-periodic profiles
of the zonal velocity. For $C=1$, the zonal profiles are almost evenly
spaced, but one can see that there is still some turbulent activity
within the large-scale flows. On the contrary, the right plot for
$C=10$ is closer to the asymptotic adiabatic regime. The zonal velocity
profile is less regular and exhibits several scales (which corresponds
to zonal flows of different sizes, as discussed in Section \ref{sec:C-dep}),
and turbulence is more strongly suppressed in this system, even though
we notice some large-scale eddies advected by the sheared flows. Another
feature of this high C state is the asymmetry of the zonal jets, with
wide negative zonal curvature (or vorticity gradient, since $\partial_{x}^{2}\overline{v}_{y}=\partial_{x}\overline{\Omega}$)
regions joined together with very narrow (but peaked) positive curvature
regions. This results in a narrowly peaked zonal velocity profile,
also common in giant planetary atmospheres \citep{scott:2012}.

The key observation here is that the system undergeoes a symmetry
breaking when $C$ is varied, resulting in the transition from 2D
isotropic turbulence to a zonal flow dominated quasi-1D state. Below we
discuss the nature of this transition.

\subsubsection{Zonal kinetic energy and enstrophy fractions \label{subsec:zonalfrac}}

\begin{figure*}[htbp]
\begin{centering}
\includegraphics[width=2\columnwidth]{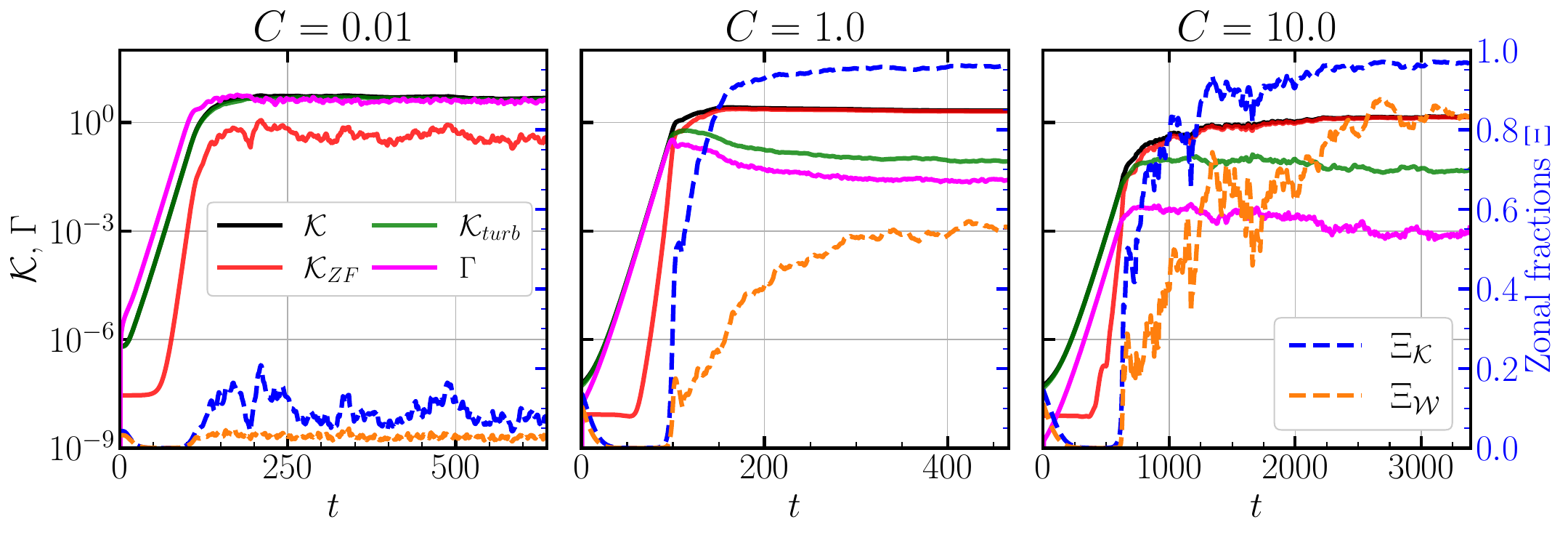}
\par\end{centering}
\caption{Time evolution of the kinetic energy and the particle flux for $C=0.01$
(left), $C=1$ (middle) and $C=10$ (right) from simulations with a
padded resolution of ${\color{teal}{\normalcolor 4096\times4096}}$.
The total kinetic energy $K$ is in black, the zonal kinetic energy
$K_{ZF}$ in red and the turbulent (\emph{i.e. }non-zonal) kinetic
energy $K_{turb}\equiv K-K_{ZF}$ in green. The particle flux is in
magenta. The zonal energy fraction $\Xi_{K}=K_{ZF}/K$ and enstrophy
fraction $\Xi_{W}=W_{ZF}/W$ correspond respectively to the blue and
orange dashed lines. \label{fig:energy_zonal_fraction}}
\end{figure*}

In order to study the formation of zonal flows, which are intrinsically
non-linear structures driven by energy or enstrophy transfer from
turbulence, we first look at the kinetic energy of these different
modes. The kinetic energy of the system averaged over all the domain can be written as
\begin{equation}
K\equiv\frac{1}{2}\langle v^{2}\rangle_{x,y}=\frac{1}{2}\sum_{k_{x},k_{y}}k^{2}|\phi_{k}|^{2}\:,\label{eq:Ktot}
\end{equation}
using $\mathbf{v}=\boldsymbol{\hat{z}}\times\nabla\phi\xrightarrow{\mathcal{F}}\mathbf{v}_{k}=i(-k_{y}\mathbf{e}_{x}+k_{x}\mathbf{e}_{y})\phi_{k}$
and Parseval's theorem. It can be decomposed into a zonal part
\begin{equation}
K_{ZF}\equiv\frac{1}{2}\langle\overline{v}_{y}^{2}\rangle_{x}=\frac{1}{2}\sum_{q_{x}}q_{x}^{2}|\overline{\phi}_{q_{x}}|^{2}\:,\label{eq:KZF}
\end{equation}
 and a non-zonal part
\begin{equation}
K_{turb}\equiv\frac{1}{2}\langle\widetilde{v}^{2}\rangle_{x,y}=\frac{1}{2}\sum_{k_{x}}\sum_{k_{y}\neq0}k^{2}|\phi_{k}|^{2}\:,\label{eq:Kturb}
\end{equation}
and we have of course $K_{ZF}+K_{turb}=K$.

In Figure \ref{fig:energy_zonal_fraction}, we show the time evolution
of the total, zonal and non-zonal kinetic energies, for $C=0.01$
(left), $C=1$ (middle) and $C=10$ (right) respectively. The initial
evolution of the fluctuation energy clearly exhibits the exponential
growth of the total and non-zonal energies (black and green lines)
due to the linear instability, quickly followed by the non-linear
growth of the zonal energy (the red line) through the nonlinear drive
and/or the modulational instability mechanisms. For simulations which
exhibit zonal flows, \emph{i.e.} $C=1$ and $C=10$, the zonal kinetic
energy $K_{ZF}$ tends to be very close to the total kinetic energy
$K$ after saturation. On the contrary, when the system is close to
the hydrodynamic regime, \emph{i.e.} $C=0.01$, the total kinetic energy
is mostly composed of the turbulent - \emph{i.e.} non-zonal - kinetic
energy.

This difference between zonal and turbulent kinetic energy levels
at saturation suggests that a relevant quantity that can be used to
characterise the preponderance of zonal flows is the ratio between
the zonal kinetic energy and the total kinetic energy, which we will
call \emph{the zonal energy fraction}:
\begin{equation}
\Xi_{K}\equiv\frac{K_{ZF}}{K}=\frac{\sum_{q_{x}}q_{x}^{2}|\overline{\phi}_{q_{x}}|^{2}}{\sum_{k_{x},k_{y}}k^{2}|\phi_{k}|^{2}}\:.\label{eq:zkf}
\end{equation}

This fraction measures the amount of energy in the zonal modes and
is frequently used to quantify the predominance of zonal flows \citep{numata:2007,grander:2024}.
It can also be used as an indicator of symmetry breaking between 2D
isotropic turbulence ($\Xi_{K}\ll1$ when there is equipartition of
energy between all Fourier modes), and a quasi-1D system, dominated
by large radial structures ($\Xi_{K}\lesssim1$). Because of this
feature of the zonal energy fraction, \emph{i.e.} that it goes from $0$
to 1 at the transition from 2D to quasi-1D as the energy injection
and hence the ``effective temperature'' is decreased, we further
speculate that it can also be used as an order parameter of the phase
transition formalism.\textcolor{magenta}{{} }Note that while the concept
of ``effective temperature'' for such a system is not a well-defined
one, even with a primitive definition based on the non-zonal kinetic
energy\emph{ i.e.} $\mathcal{T}\propto(K-K_{ZF})$ one can see that
the quasi-1D zonostrophic turbulence state is substantially ``colder''
than the eddy dominated 2D turbulence state.

Likewise, we define the ``\emph{zonal enstrophy fraction''} as
\begin{equation}
\Xi_{W}\equiv\frac{W_{ZF}}{W}=\frac{\sum_{q_{x}}q_{x}^{4}|\overline{\phi}_{q_{x}}|^{2}}{\sum_{k_{x},k_{y}}k^{4}|\phi_{k}|^{2}}\:.\label{eq:zwf}
\end{equation}

Note that one can also use the fractions based on total energy $n^{2}+|\nabla\phi|^{2}$
and potential enstrophy $\left(n-\nabla^{2}\phi\right)^{2}$, but
since our goal is to characterise zonal flows, focusing on flows and
using kinetic energy and enstrophy fractions appear as natural choices.
It also has the advantage of abstraction, in the sense that one can
apply it directly to a different system as long as the nonlinear flow
evolution can be written as an advection of vorticity.

The zonal energy fraction $\Xi_{K}$ and enstrophy fraction $\Xi_{W}$
are shown in Figure \ref{fig:energy_zonal_fraction}, respectively
as dashed blue and orange lines. While about $10\%$ of the kinetic
energy is stored in the zonal modes for $C=0.01$, the zonal energy
fraction approaches $100\%$ for $C=1$ and $C=10$, where zonal flows
emerge and dominate the system. A similar behaviour is observed for
the zonal enstrophy, although it reaches only $60\%$ and $80\%$
of the total enstrophy, respectively for $C=1$ and $C=10$, and fluctuates
more in time. This can be explained by the fact that eddies are still
present in simulations where zonal flows are developed, as seen on
Figure \ref{fig:Snapshots-of-vorticty}, and they contribute more
to enstrophy.

\subsubsection{Radial particle flux}

Another important observable of the system is the radial particle
flux due to the fluctuating $E\times B$ flow, corresponding to turbulent
particle transport. One of the goals of studying this system is to
construct reduced models that can estimate transport levels, self
consistently with zonal flows. The turbulent radial particle flux,
averaged over the entire 2D space, is defined as
\begin{equation}
\Gamma\equiv\langle\widetilde{n}\widetilde{v}_{x}\rangle_{x,y}\,,\label{eq:part-flux}
\end{equation}
 where $\widetilde{n}$ and $\widetilde{v}_{x}$ are the fluctuating
density and the radial $E\times B$ velocity respectively, and $\langle\cdot\rangle_{x,y}$
denotes averaging over both $x$ and $y$ directions. The time evolution
of the particle flux for $C=0.01,$ 1 and 10 is shown in Figure \ref{fig:energy_zonal_fraction}
in magenta. Using the Fourier decomposition of both fields, we can
re-write this as
\begin{equation}
\Gamma=\text{Re}\left[\sum_{k_{x},k_{y}}ik_{y}n_{k}\phi_{k}^{*}\right]=-\sum_{k_{x},k_{y}}k_{y}\text{Im}\left[n_{k}\phi_{k}^{*}\right]\,.\label{eq:part-flux-four}
\end{equation}

The simplest way to estimate this quantity is to use the \emph{quasi-linear
}flux, derived using the linear relation between $\phi_{k}$ and $n_{k}$
(which is only valid in the linear regime but extrapolated in the
quasi-linear approach) \citep{kadomtsev:book:plasturb,bourdelle:07,stephens:2021}.
However, the turbulent spectral intensity $|\phi_{k}|^{2}$ remains
unknown in this formulation. In order to predict the particle transport,
especially in gyrokinetic transport codes, an additional assumption
on $|\phi_{k}|^{2}$ is sometimes provided through the so-called \emph{``saturation
rule''} (see Refs. \onlinecite{bourdelle:07,casati:2012,parker:2023} for example),
which relies on the mixing length assumption to determine the level
of fluctuations, and eventually the particle flux.

Here, in order to apply this concept, consider the following equation
for the non-zonal fluctuation level $\mathcal{E}_{k}$ (representing
for example the non-zonal potential enstophy), near its peak: 
\begin{equation}
\partial_{t}\mathcal{E}_{k}=\gamma_{k}\mathcal{E}_{k}-D\left(\mathcal{E}\right)k_{y}^{2}\mathcal{E}_{k}\,,\label{eq:balance}
\end{equation}
where $\gamma_{k}\mathcal{E}_{k}$ represents linear injection and
$D\left(\mathcal{E}\right)k_{y}^{2}$ represents the nonlinear transfer
away from the peak, as a function of the total fluctuation level $\mathcal{E}$.
Using renormalisation \emph{à la }Dupree in the form of mixing-length
\citep{dupree:1967,krommes:2002,weiland:2016}, we can identify $D(\mathcal{E})$
as being the diffusion coefficient, which also appears in the flux-gradient
relation: $\Gamma\approx-D_{turb}\nabla n_{0}\approx\kappa D_{turb}$.
This coefficient can be estimated assuming steady state in the peak
$k_{y}$ of (\ref{eq:balance}):
\[
D_{turb}\sim D\left(\mathcal{E}\right)\sim\frac{\gamma_{k_{y}}}{k_{y}^{2}}\biggl|_{max}\,,
\]
taking the maximum value of the ratio $\frac{\gamma_{k_{y}}}{k_{y}^{2}}$
which contains the chacteristic time and length scales of the instability-driven
turbulence. When this is true, we can write the saturated flux as
\begin{equation}
\Gamma_{sat}=\kappa\sum_{k_{y}}S(k_{y})\frac{\gamma_{k_{y}}}{k_{y}^{2}}\biggl|_{max}\,,\label{eq:gamma-sat}
\end{equation}
where $S(k_{y})$ is a model $k_{y}$ spectrum, which peaks the spectral
intensity around the maximum value of $\frac{\gamma_{k_{y}}}{k_{y}^{2}}$
with a slope of $k^{-3}$ (see Ref. \onlinecite{casati:2012} for details).
Note that the summation is carried only along the ``poloidal'' (\emph{i.e}
the\emph{ }$y$) axis.

However, this expression does not account for the damping effect of
zonal flows on the particle transport. This is a consequence of taking
$\mathcal{E}\rightarrow\mathcal{E}_{k}$ implicitly in the $\mathcal{E}$
dependence of $D\left(\mathcal{E}\right)$ when going to $D_{turb}$.
However, before reaching the steady state, part of the energy that
has been injected in the system was transfered to zonal flows, which
do not generate any radial transport. Therefore, assuming $D\left(\mathcal{E}\right)\propto\mathcal{E}$,
we can write 
\[
D_{turb}\sim\left(1-\Xi_{K}\right)D\left(\mathcal{E}\right)\sim\left(1-\Xi_{K}\right)\frac{\gamma_{k_{y}}}{k_{y}^{2}}\biggl|_{max}\,,
\]
which allows us to represent the reduction of the saturated flux due
to zonal flows, \emph{via} the fraction of energy in the turbulent modes,
written as $1-\Xi_{K}$. Indeed, since the saturation rule estimates
the level of turbulence and the zonal flows generate zero radial transport,
the $1-\Xi_{K}$ factor is a way of accounting for how much of the
system's energy is dedicated to the radial transport (\emph{i.e.}
the velocity in the $x$ direction). Note that it is somewhat more
common to use the zonal shearing $v_{E}'$ and divide the flux by $1+\alpha v_{E}'{}^{2}$,
especially in transport models, to account for the effect of sheared
poloidal flows \citep{hinton:93,miki:12,bufferand:16}. But, as we
will show below in Section \ref{subsec:Power-law-scalings}, our proposition
matches the numerical observations rather well.

\section{Transition from 2D isotropic turbulence to a quasi-1D system and
hysteresis behaviour\label{sec:C-dep} \label{sec:C-dep}}

We describe and analyse the transition from 2D isotropic turbulence
to 1D zonal flows that we observe to occur around $C\approx0.1$,
using numerical simulations of the Hasegawa-Wakatani equations accross
a range $C\in[10^{-4},20]$. From a phase transition point of view,
this is akin to the transition from a hotter gas to a colder liquid
or from 3D to 2D turbulence. Such transition may exhibit hysteresis
behaviour around the critical point. In the case of a phase transition
in materials (\emph{e.g.} from solid to liquid), the hysteresis is
a consequence of the fact that the breaking of the organised state,
such as a solid, requires some energy absorption, or \textit{latent heat},
suggesting a first order transition. In our system, we observe a similar
hysteresis at the transition point by looking at the two order parameters
we defined in Part \ref{subsec:zonalfrac}, \emph{i.e. }the zonal
energy and enstrophy fractions.

\subsection{Simulation details\label{subsec:simdet}}

A large number of lower resolution simulations with a padded resolution
of $512\times512$ are performed with an extensive scan of the coupling
parameter over $C\in[10^{-4},20],$ while keeping $\kappa=1$, using
a pseudo-spectral solver with 2/3 rule for dealiasing. As shown in
the bottom panel of Figure \ref{fig:Maximum-linear-growth}, the scale
at which the energy is injected, corresponding roughly to the inverse
of the wave-number $k_{y0}$ at which the growth rate is maximum,
becomes very large for low values of $C$. Therefore, we choose to
change the box size such that it scales together with the linear injection
scale, by taking $L_{x}=L_{y}=2\pi\times(k_{y0}/10)^{-1}$, which
gives a wave-number resolution $\Delta k=k_{y0}/10$. The goal of
such scaling is to ensure correct energy injection for each value
of $C$. Otherwise, we would truncate it for low injection scales
with a too small box, which would result in scaling problems when
changing $C$. In order to avoid numerical issues or considering too
much dissipation, the diffusion coefficients are tuned to ensure that
the energy injection $\gamma_{max}$ is balanced by the small-scale
dissipation $\nu k_{m}^{2}$, where $k_{m}$ is the largest wave-number
of the spectral grid. In practice, we use 
\begin{equation}
\nu(C)=D(C)=0.017\times\frac{\gamma_{max}}{k_{y0}^{2}}\,,\label{eq:nu_dep}
\end{equation}
where the numerical prefactor $0.017$ was set empirically. For $C=1$,
we have $\nu\approx1.1\times10^{-3}.$

The initial condition is chosen as an isotropic Gaussian \emph{seed}
in Fourier space of maximum amplitude $A=10^{-4}$ centered on $(k_{x},k_{y})=(0,0)$
and with a standard deviation $\sigma_{k_{x}}=\sigma_{k_{y}}=0.5$,
and random initial phases. Each simulation is run until $t=100\times\gamma_{max}^{-1}$,
which is found to be enough to observe the system for a sufficient
time in the saturated state.

We also compare the results with data from the high resolution simulations
presented in Part \ref{subsec:Non-linear-behaviour:-hydrodynam},
the details of which are given in Table \ref{tab:4096_params}.

\subsection{Features of the 2D-1D transition}

In this part, we present the aspects of the transition that we observed,
using mainly the quantities that we have previously defined in Section
(\ref{sec:The-Hasegawa-Wakatani-equations}).

\subsubsection{Zonal velocity profiles}

\paragraph{Time evolution of the zonal velocity profiles}

\begin{figure*}[htbp]
\begin{centering}
\includegraphics[width=2\columnwidth]{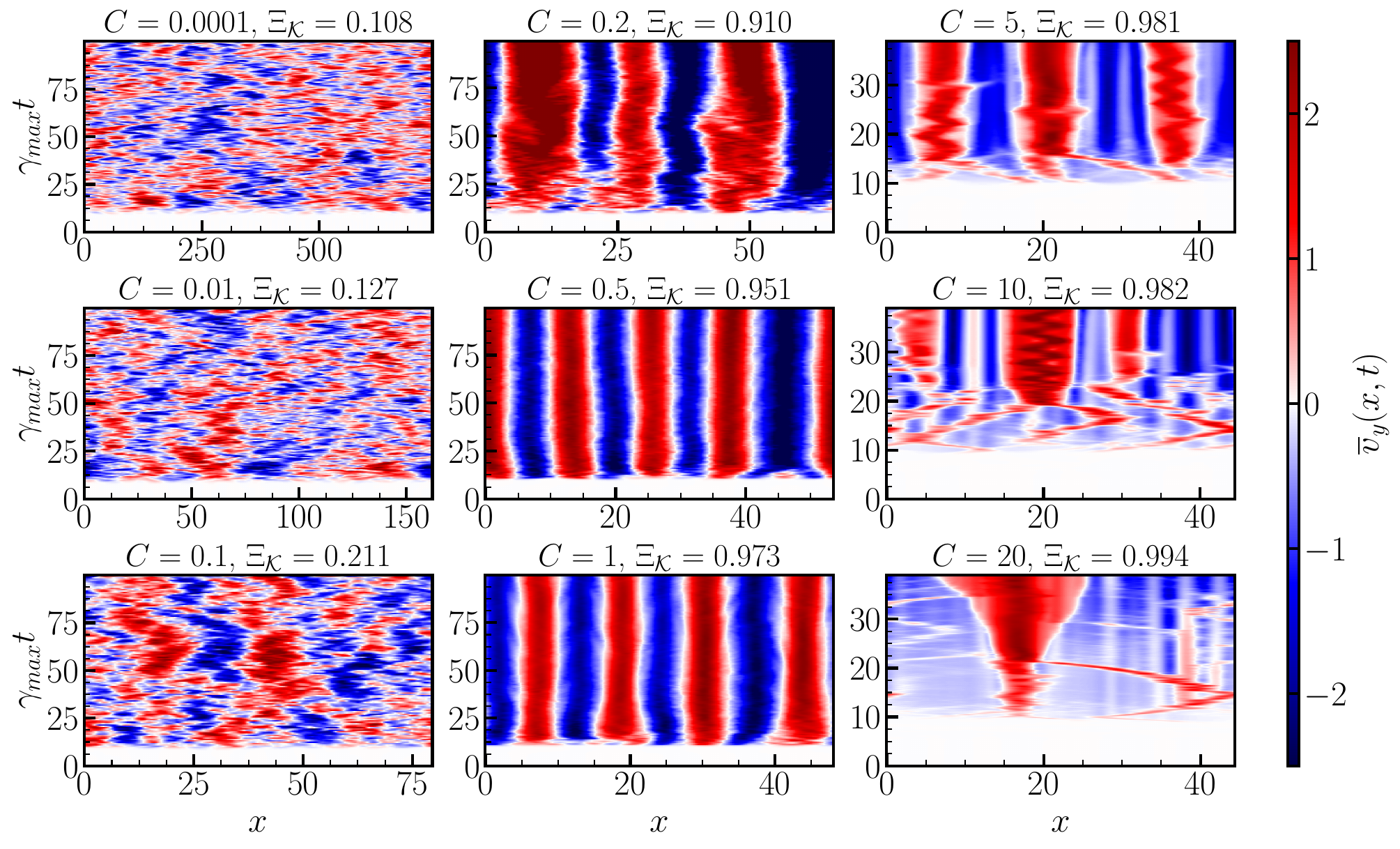}
\par\end{centering}
\caption{Zonal velocity profiles for different values of C, as a function of
the radial coordinate ($x$-axis) and time ($y$-axis). The time is
normalised to the maximum growth rate for each value of $C$. The
value of $C$ and the zonal kinetic energy fraction $\Xi_{K}$, averaged
over the final quarter of the simulation, are given for each plot
in the title. Note that for the right column, we show the profiles
only up to $\gamma_{max}t\approx40$ in order to highlight the merging
of zonal flows which is mainly observed in the early formation of
the profiles.\textcolor{red}{{} }\label{fig:Zonal-velocity-profiles}}
\end{figure*}

As a first demonstration of the transition between a 2D turbulent
state and a 1D zonal flow dominated regime, we consider several profiles
of the zonal velocity $\overline{v}_{y}(x,t)\equiv\partial_{x}\overline{\phi}$
as a function of time, for different values of C in Figure \ref{fig:Zonal-velocity-profiles}.
For each value of $C$, the time is normalised by the maximum growth
rate ($t\rightarrow\gamma_{max}t)$ in order to bring the time evolution
of the different profiles to the same range. The white regions at
the bottom of each plot corresponds to the initial linear growth.
The value of the zonal kinetic energy fraction averaged over the final
quarter of the simulation is given at the top of each plot. One can
clearly observe three different categories of evolutions for the profiles
depending on the value of C:

\begin{enumerate}
\item \textbf{Chaotic, disordered and random profiles}: these profiles are
associated with $C\leq0.1$ and do not display any particular regularity
in the radial direction (left column of Figure \ref{fig:Zonal-velocity-profiles}).
They correspond to the turbulent state, in which the system features
eddies randomly moving accross the 2D plane. The low values of the
zonal energy fraction $\Xi_{K}$ obtained for these simulations witness
that the system is dominated by turbulence, even though the fraction
remains non-zero. For $C=0.1$, we start to see the emergence of large-scale
zonal patterns.
\item \textbf{Emergence of radially structured stationary profiles}: for
$C>0.1$, we see the emergence of long-lived quasi-periodic radial
structures as $C$ increases (middle column of Figure \ref{fig:Zonal-velocity-profiles}).
Although somewhat noisy and blurry for $C\gtrsim0.1$, the zonal velocity
profiles become more and more steady as $C$ approaches 0.5, which
is associated with a sudden jump of the zonal energy fraction to high
values close to 1. For $C\in[0.5,3]$, the system displays quite steady
and regular zonal patterns. The zonal velocity profiles are smooth
and the position of their peaks doesn't move.\textcolor{magenta}{{}
}The number of positive peaks slightly changes from 2 to 5 between
the different values of $C$ (note that the length of the box decreases
with increasing $C$).
\item \textbf{Meandering zonal flows}: for high values of $C$, especially
for $C>3$, the system starts to exhibit zonal flows moving radially
at a roughly constant speed and merging with larger stationary structures
(right column of Figure \ref{fig:Zonal-velocity-profiles}). In this
state, zonal velocity profiles are less regular, and feature a wider
range of radial scales, with large zonal flows modulating smaller
ones, which makes it harder to define a single characteristic scale.
For these large values of $C$ the system remains strongly zonostrophic,
with zonal energy fractions close to 1. Apart from the fact that forcing
is still due to an instability, albeit very weak, this state is very
close to the adiabatic regime, and therefore is governed by the Charney-Hasegawa-Mima
equation at leading order. In simulations of forced $\beta$-plane
turbulence, similar meandering and merging of zonal jets have been
observed \citep{cope:2021}, and possibly attributed to the presence
of solitary nonlinear waves called zonons \citep{sukoriansky:2008}.
As simulations at high-$C$ values require very long times to run
and correspond to a limit of the Hasegawa-Wakatani equations with
a very weak linear instability (see Figure \ref{fig:Maximum-linear-growth}),
a detailed study of the system in this limit is out of the scope of
the present article.
\end{enumerate}

\paragraph{Characteristic wave-number of zonal flows}

\begin{figure}[htbp]
\begin{centering}
\includegraphics[width=1\columnwidth]{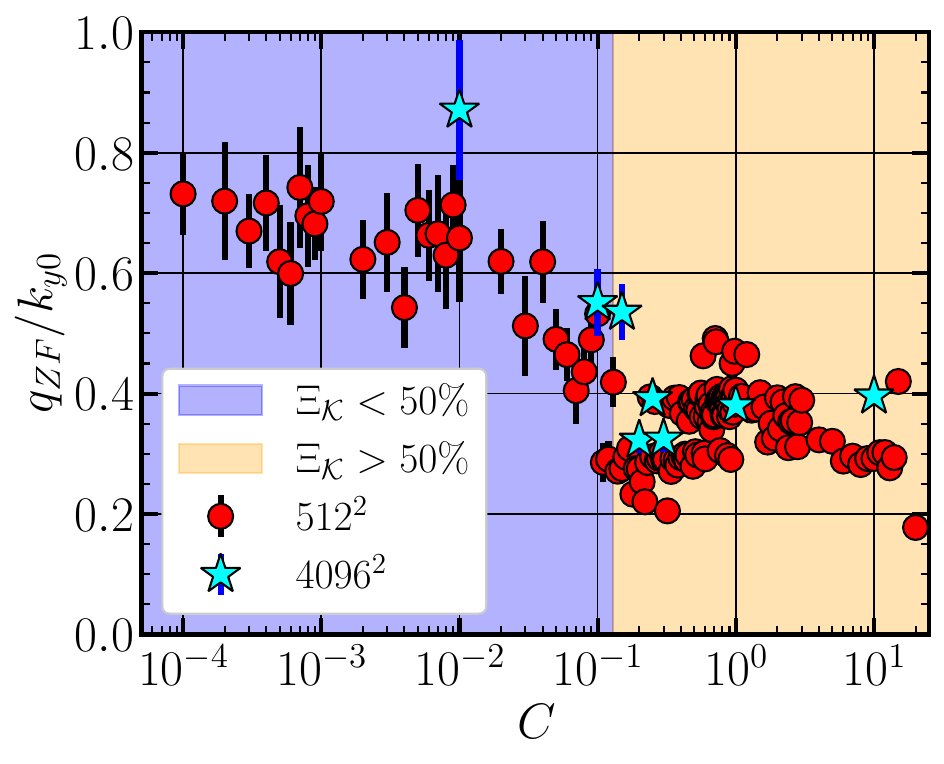}
\par\end{centering}
\caption{Ratio of the dominant zonal wave-number to the injection wave-number
$q_{ZF}/k_{y0}$, averaged over $t$ in the final quarter of each
simulation, versus $C$. Red dots correspond to the extensive scan
performed with the low resolution (i.e. $512^{2}$) simulations, while
the light blue stars correspond to the high resolution runs ($4096^{2})$
used in Section \ref{subsec:Non-linear-behaviour:-hydrodynam}. Blue
and yellow regions indicate respectively when the zonal energy fraction
$\Xi_{K}$ is smaller (eddy dominated state) or larger (zonal flow
dominated state) than 50\%. \label{fig:qZF_C}}
\end{figure}

Another interesting measure is the dependency of the characteristic
wave-number - or the radial size - of the zonal flows to the adiabaticity
parameter $C$. The dominant zonal wave-number can be defined using
the average radial wave-number weighted by the zonal kinetic energy
at that wave-number\textcolor{blue}{{} \citep{numata:2007}}: 
\begin{equation}
q_{ZF}=\frac{\sum_{q_{x}}|q_{x}|E_{q_{x}}}{\sum_{q_{x}}E_{q_{x}}}\:,\label{eq:qZF}
\end{equation}
where $E_{q_{x}}\equiv q_{x}^{2}|\overline{\phi}_{q_{x}}|^{2}$ is
the zonal kinetic energy at the wave-number $q_{x}$. In Figure \ref{fig:qZF_C},
we show the ratio between the characteristic zonal wave-number computed
in our simulations divided by the injection wave-number proxy, namely
$q_{ZF}/k_{y0}$, as a function of the adiabaticity parameter $C$.
In this way, we can visualise how the ratio of the typical zonal width
to the injection scale changes as $C$ is varied. We also show the
extent of the turbulent and zonostrophic states, respectively the
blue and yellow regions, using the zonal energy fraction. The former
is dominated by turbulence and corresponds to $\Xi_{K}<50\%$, while
the latter is dominated by zonal flows and is associated with $\Xi_{K}>50\%$.

One can see that the dominant zonal mode is always smaller than the
injection wave-number, which is loosely consistent with the general
picture of the inverse cascade forming large-scale structures. However,
the relation between $q_{ZF}$ and $k_{y0}$ seems to be different
in the two states.

In the small $C$ regime (blue region), the dominant zonal wave-number
is slightly smaller than the most unstable linear mode and the values
of the ratio $q_{ZF}/k_{y0}$ seem to fluctuate from one value of
$C$ to another. This can be explained by the fact that this regime
is dominated by chaotic eddies which have the same size in both dimensions
since they are rotating. The fact that this size is larger than the
injection scale is linked to a standard observation that the turbulent
energy spectrum is usually maximum at a wave-number smaller than the
most unstable mode, which shifts the energy peak towards larger scales.
Close to the transition point, the zonal wave-number becomes significantly
smaller than the most unstable linear mode, which indicates that the
system starts to favor larger radial scales.

On the contrary, in the regime dominated by zonal flows, the ratio
$q_{ZF}/k_{y0}$ is roughly piecewise constant as $C$ is varied,
wich means that the characteristic zonal wave-number scales linearly
with $k_{y0}$, although the linear coefficient changes with $C$
and sometimes two (or more) values of the ratio seems to coexist.
For $C\in[0.1,1]$, $q_{ZF}$ scales like $0.3k_{y0}$, then for $C\in[0.5,3]$,
we have mainly $q_{ZF}\approx0.4k_{y0}$ (although the linear factor
reaches 0.5 in a few simulations). For $C>3$, we have again $q_{ZF}\approx0.3k_{y0}$.
Note that this observation is somewhat biased, since we choose the
wave-number grid resolution to be $\Delta k=k_{y0}/10$, so the available
radial wave-numbers are basically $q_{xn}=n\times0.1k_{y0}$. Since
zonal flows are not purely monochromatic structures, their true kinetic
energy spectrum can be quite flat around the maximum. As only few
modes are available for the large scales, the maximum can be randomly
distributed among these modes, which can explain the superimposition
of multiple $q_{xn}$. The decrease of the zonal wave-number for large
values of $C$ is consistent with the merging and the modulation of
zonal flows by larger ones observed in Figure \ref{fig:qZF_C}.

Note also that the highly resolved simulations follow roughly the
same evolution with $C$.

\subsubsection{Transition in zonal energy and enstrophy fractions}

\begin{figure*}[htbp]
\begin{centering}
\includegraphics[width=0.7\paperwidth]{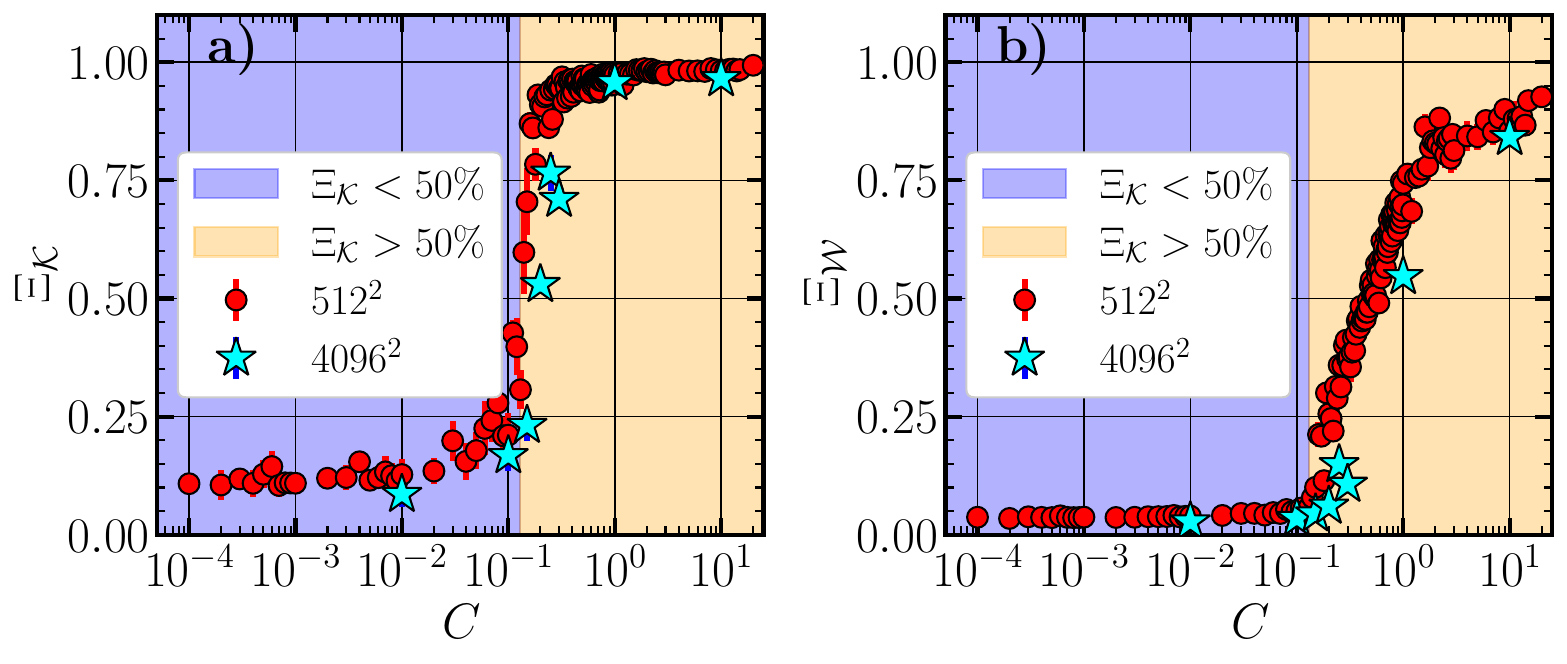}
\par\end{centering}
\caption{Zonal kinetic energy fraction $\Xi_{K}$ (a) and zonal enstrophy fraction
$\Xi_{W}$ (b) as functions of the adiabaticity parameter $C.$ Mean
value and standard deviation of the zonal fractions have been computed
by averaging over $t$ in the final quarter of each simulation. Red
dots correspond to the low resolution extensive scans (\emph{i.e.} $512^{2}$).
Light blue stars correspond to the high resolution simulations ($4096^{2})$
shown in Part \ref{subsec:Non-linear-behaviour:-hydrodynam}. Blue
and yellow regions correspond to $\Xi_{K}<50\%$ and $\Xi_{K}>50\%$,
respectively. The result for the zonal energy fraction is very similar
to those of Refs. \onlinecite{numata:2007,grander:2024}. \label{fig:Zonal-fraction-C}}
\end{figure*}

The observation of the zonal velocity profiles for different values
of $C$ suggests that the transition from the turbulent regime to
a state dominated by zonal flows occurs at around $C\approx0.1$.
This can also be measured by looking at the zonal energy and enstrophy
fractions as a function of $C$. In Figure \ref{fig:Zonal-fraction-C},
the mean value and the standard deviation of the zonal energy fraction
$\Xi_{K}$ (left plot), and of the zonal enstrophy fraction $\Xi_{W}$
(right plot), both computed over the final quarter of each simulation,
are shown (red circles and light blue stars correspond respectively
to $512^{2}$ and $4096^{2}$ simulations).

The dependency of the zonal energy fraction to $C$ clearly evidences
the transition between a 2D turbulent system ($\Xi_{K}<50\%$, blue
region in both plots) and a regime dominated by 1D zonal flows for
high values of $C$ ($\Xi_{K}>50\%$, yellow region in both plots)
at around $C\approx0.1$. This result is very similar to those of
Refs. \onlinecite{numata:2007,grander:2024}. Furthermore, scans in $\kappa$
with fixed $C$ were performed in Ref. \onlinecite{numata:2007}, yielding
the same transition point $C/\kappa\approx0.1$. The zonal fraction
exhibits a sudden jump from $10\%$ in the turbulent state to almost
$100\%$ in the zonal dominated regime, which justifies the use of this fraction
as an order parameter for the transition. Note that in the eddy dominated
state, the zonal fraction is not $0\%$, since the energy is roughly isotropically
distributed. With increasing resolution, the fraction should reach
smaller values in this state, because the ``weight'' of the $x$-axis
compared to the other directions is decreased. Note also that the
data around the transition point exhibits larger standard deviations.
These points correspond to simulations within the transition region,
where the system can either develop zonal flows or stay in the eddy
dominated state, somewhat analogous to multi-phase flows.

The zonal enstrophy fraction $\Xi_{W}$ exhibits a similar transition
from less than $5\%$ (low-$C$) to approximately $80\%$, suggesting that
it may reach unity asymptotically. However, in contrast to the zonal
energy fraction, which makes a sudden jump around $C\approx0.1$,
the zonal enstrophy fraction remains at low values before the transition
point and then grows progressively as $C$ is increased. As previously
discussed in Part \ref{subsec:zonalfrac}, the enstrophy fraction
reaches smaller values than the energy fraction in the zonal dominated
regime because small scale turbulence is still present within the
zonal flows, especially for $C\in[0.1,1]$. Since 2D turbulence features
a direct enstrophy cascade and the enstrophy of the mode $k$ is $W_{k}=k^{4}|\phi_{k}|^{2}$,
as opposed to $E_{k}=k^{2}|\phi_{k}|^{2}$, the small turbulent scales
contribute more to the total enstrophy. And since there are a large number
of eddies close to the transition that are gradually suppressed when
$C$ increases (see Figure \ref{fig:Snapshots-of-vorticty} middle
and right plots), the zonal enstrophy fraction can only slightly increase
at the transition point. On the contrary, the energy is associated
with large scale flows, since the large scale zonal velocity tends
to be larger than the velocity associated with the small eddies, which
explains why the zonal energy fraction suddenly jumps when zonal flows
start to form at the transition.

Note that for both fractions, the results from high resolution simulations
are very similar to the low resolution ones, although slightly shifted
towards higher $C$.

In simulations with the same padded resolution ($512^{2}$) but with
higher dissipation ($3$ times the value $\nu(C)$ given by \ref{eq:nu_dep}),
we noticed that the transition was shifted towards smaller $C$ (not
shown here). This may indicate that the dissipation scale plays some
minor role in setting the exact $C$ value at which the system transitions.
This dependency, which needs to be investigated more carefully, has
also been observed in the minimal reduced model that is able to reproduce
the transition, which is discussed in Section \ref{sec:Reductions},
suggesting that such a reduced model may actually be used in order
to understand the role of dissipation in this transition. On the other
hand, lowering the dissipation in the high resolution simulations
(we try deviding by 5 the values in Table \ref{tab:4096_params})
does not seem to shift the transition towards higher $C$, and the
results (not shown here) were quite similar to the light blue stars
in Figure \ref{fig:Zonal-fraction-C}.

We also performed simulations (not shown here) with a domain that
is $4$ times larger and with a resolution of $2048^{2}$, so that
the smallest scale available is the same as the original extensive
scan with resolution $512^{2}$, in order to approach the ``ideal
thermodynamic state'', where we observed a sharpening of the transition
for the zonal energy fraction, again supporting the hypothesis of
a 1\textsuperscript{st} order phase transition.

Finally, note that zonal fractions of total energy (instead of kinetic
energy) and potential enstrophy (instead of enstrophy) also display
very similar transitions, which we don't show here to avoid overloading
the discussion.

\subsubsection{Power law scalings for the radial particle flux\label{subsec:Power-law-scalings}}

\begin{figure}[htbp]
\begin{centering}
\includegraphics[width=1\columnwidth]{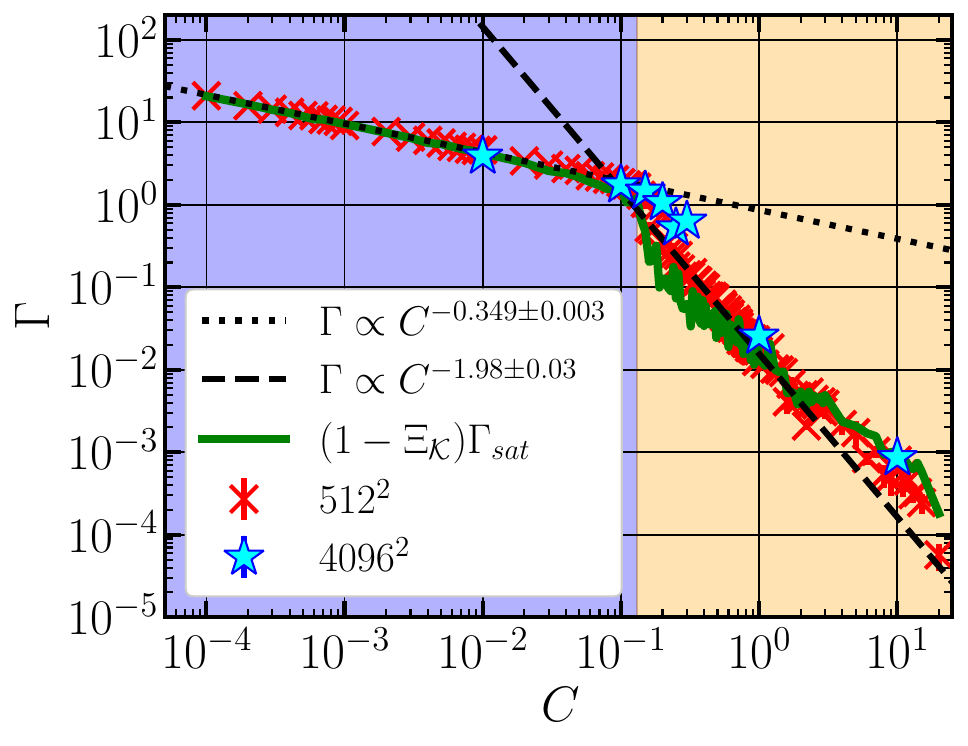}
\par\end{centering}
\caption{Radial particle flux (red crosses: $512^{2}$ resolution, light blue
stars: $4096^{2}$ resolution) as a function of $C$, compared with
the saturation rule formulation (green) given by (\ref{eq:gamma-sat})
and multiplied by the turbulent energy fraction $1-\Xi_{K}$. Values
are averaged over the final quarter of each simulation duration. The
flux is fit by two power laws: $\Gamma\propto C^{-0.35}$ for $C\in[10^{-4},1.3\times10^{-1}]$
(dotted) and $\Gamma\propto C^{-2}$ for $C\in[1.3\times10^{-1},2]$
(dashed). Blue and yellow regions correspond respectively to $\Xi_{K}<50\%$
and $\Xi_{K}>50\%$. \label{fig:FLUX-C}}
\end{figure}

The transition can also be observed in the radial particle flux $\Gamma$.
In Figure \ref{fig:FLUX-C}, we show the particle flux, averaged over
the final quarter of each simulation (red crosses and light blue stars
correspond respectively to $512^{2}$ and $4096^{2}$ runs), where
two different power law dependencies to the adiabaticity parameter
$C$ between low-$C$ and high-$C$ regimes can be observed. In the
low-$C$ branch ($C\in[10^{-4},1.3\times10^{-1}]$), we find $\Gamma\propto C^{-0.35}$
(dotted line), and in the high-$C$ branch ($C\in[1.3\times10^{-1},2]$),
we find $\Gamma\propto C^{-2}$ (dashed line). The transition between
the two power laws seems to occur at the point of transition of the
zonal energy fraction, around $C\approx0.1$ (as shown by the blue
and yellow regions). For $C>2$, the flux seems to depart from
the power law and follows a less steep dependency. However, since
the flux becomes very intermittent due to meandering and merging zonal
flows in this regime, and the long simulations are relatively expensive,
the details of this behaviour is left to a future study. Nevertheless,
simulations with $512^{2}$ resolution but higher dissipation seem
to depart less from the power law in the high $C$ regime. In this
limit, the linear time is dominated by the real frequency, which tends
asymptotically towards the Hasegawa-Mima drift-wave frequency $\omega_{k,r}\approx\omega_{k}^{HM}=\frac{\kappa k_{y}}{1+k^{2}}$,
whereas the growth rate decays as $\gamma_{k}\propto\frac{1}{C}$.
Therefore, it could be more accurate to use a stronger dissipation
in this regime, which would scale as $\omega_{k,r}/k^{2}$. Note that
these scalings for the particle flux are exactly the same as the asymptotic
scalings from Ref. \onlinecite{hu:1997}, where they found these power
laws in the case of the non-modified Hasegawa-Wakatani system, but
in the asymptotic limits $C\to0$ and $C\to +\infty$. In our case,
the scaling is not confined to these limits, but is extended everywhere
except very close to the transition point.

Using a formulation of the flux based on the saturation rule (\ref{eq:gamma-sat})
multiplied by the turbulent energy fraction $1-\Xi_{K}$ (green line
in Figure \ref{fig:FLUX-C}), we find relatively good qualitative
and quantitative agreements with the full nonlinear flux. In the eddy
dominated regime, where $1-\Xi_{K}\approx1$, the saturation rule
formulation is very close to the results from the simulations. Using
the analytical expression for $\gamma_{k}$ when $C\ll1$, it is possible
to show that the saturated flux scales as $C^{-1/3}$, which is close
to the measured power law of the flux in this regime. In the zonal
flow dominated regime, the agreement between the estimated flux and
the simulations remains correct, although we loose the power law scaling
for $C\in[1.3\times10^{-1},2].$ If we remove the $1-\Xi_{K}$ factor,
the saturation rule overestimates the flux by more than one order
of magnitude factor (not shown here). Using an analytical expression
or a reduced model for $\Xi_{K}$, our \emph{ad hoc} formulation could
improve the prediction of the saturation rule in gyrokinetic simulations.
At least, $\Xi_{K}$ can be included as an additional parameter of
the model. \\ \\

\begin{figure}[htbp]
\begin{centering}
\includegraphics[width=1\columnwidth]{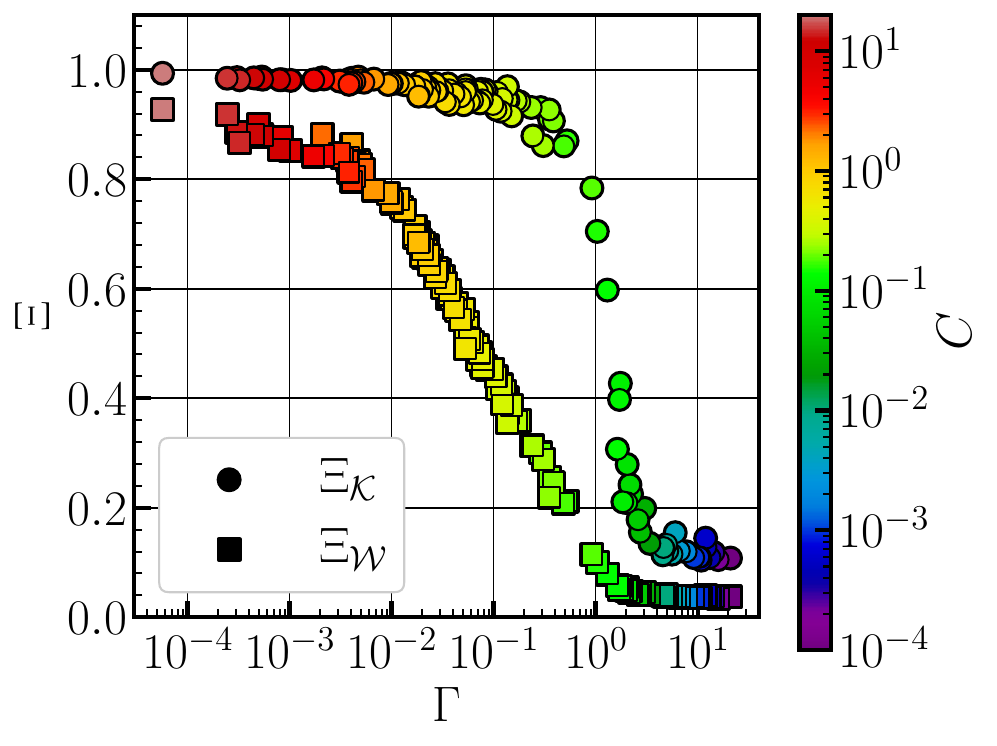}
\par\end{centering}
\caption{Zonal energy $\Xi_{K}$ (circles) and enstrophy $\Xi_{W}$ (squares)
fractions as functions of the radial particle flux $\Gamma$. The
colour scale corresponds to the value of the adiabaticity parameter
$C$ of the corresponding data points. Low values of the flux correspond
to a high level of zonal flows, while increasing $\Gamma$ leads to
2D isotropic turbulence. \label{fig:ZF_FLUX}}
\end{figure}

In order to relate the particle flux to the level of zonal flows,
we combine the previous plots in Figure \ref{fig:ZF_FLUX} and show
the zonal energy and enstrophy fractions (respectively circles and
squares) as a function of the turbulent flux $\Gamma$. The goal of
such representation is to switch from a ``gradient-driven'' perspective,
where the system evolved on constrained, \emph{a priori} fixed gradients,
to a ``flux-driven'' perspective, where the fluxes are now the input
parameters, and the gradients are the results. Although our model
belongs to the first kind, where it is $C/\kappa$ which determines
the value of the particle flux and the level of zonal flows, we can assume
that there is an underlying actual physical relation between $\Gamma$,
$C$, $\kappa$ and the zonal fraction, and switch to a perspective
where $\Gamma$ appears as a control parameter which gives us a particular
zonal fraction. This may help to make the connection to flux-driven
models, especially in turbulent transport models where density or
temperature gradients evolve as a response to input particle and heat
flux coming from the tokamak core \citep{sarazin:2021,gillot:2023},
and the scale separation between turbulence and transport is relaxed.
From this point of view, the scatter plots in Figure \ref{fig:ZF_FLUX}
suggest that low values of the particle flux would result in an ordered
state with a high level of zonal flows, while increasing $\Gamma$
would lead to more disordered 2D hydrodynamic turbulence. This representation
shows some clear and simple dependency of the zonal fractions on the
flux. It would be interesting to compare this picture to actual results
from models coupling transport and turbulence where $C/\kappa$ is
a dynamical variable wich evolves as a response to the input particle
flux $\Gamma$.

\subsection{Hysteresis in the 2D-1D transition}

\begin{figure*}[htbp]
\begin{centering}
\includegraphics[width=2\columnwidth]{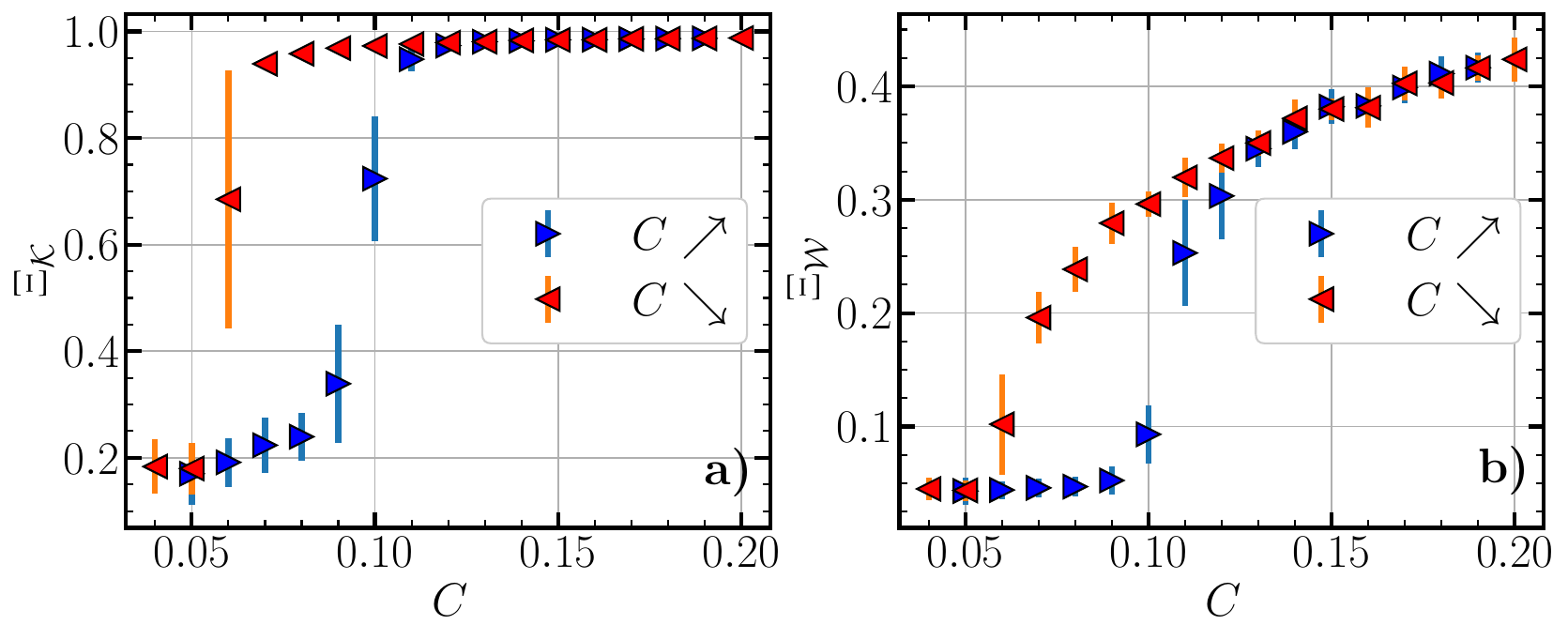}
\par\end{centering}
\caption{Hysteresis of the zonal fractions as functions of the adiabaticity
parameter $C$: a) zonal kinetic energy fraction $\Xi_{K}$, b) Zonal
enstrophy fraction $\Xi_{W}$. Blue triangles correspond to $C=0.05\rightarrow0.2$
and red triangles to $C=0.2\rightarrow0.04$. Mean values and standard
deviations of the zonal fractions are computed over $\Delta t=200\times\gamma_{max}^{-1}$
for each value of $C$. \label{fig:Zonal-faction-hyst}}
\end{figure*}

\begin{figure}[htbp]
\begin{centering}
\includegraphics[width=1\columnwidth]{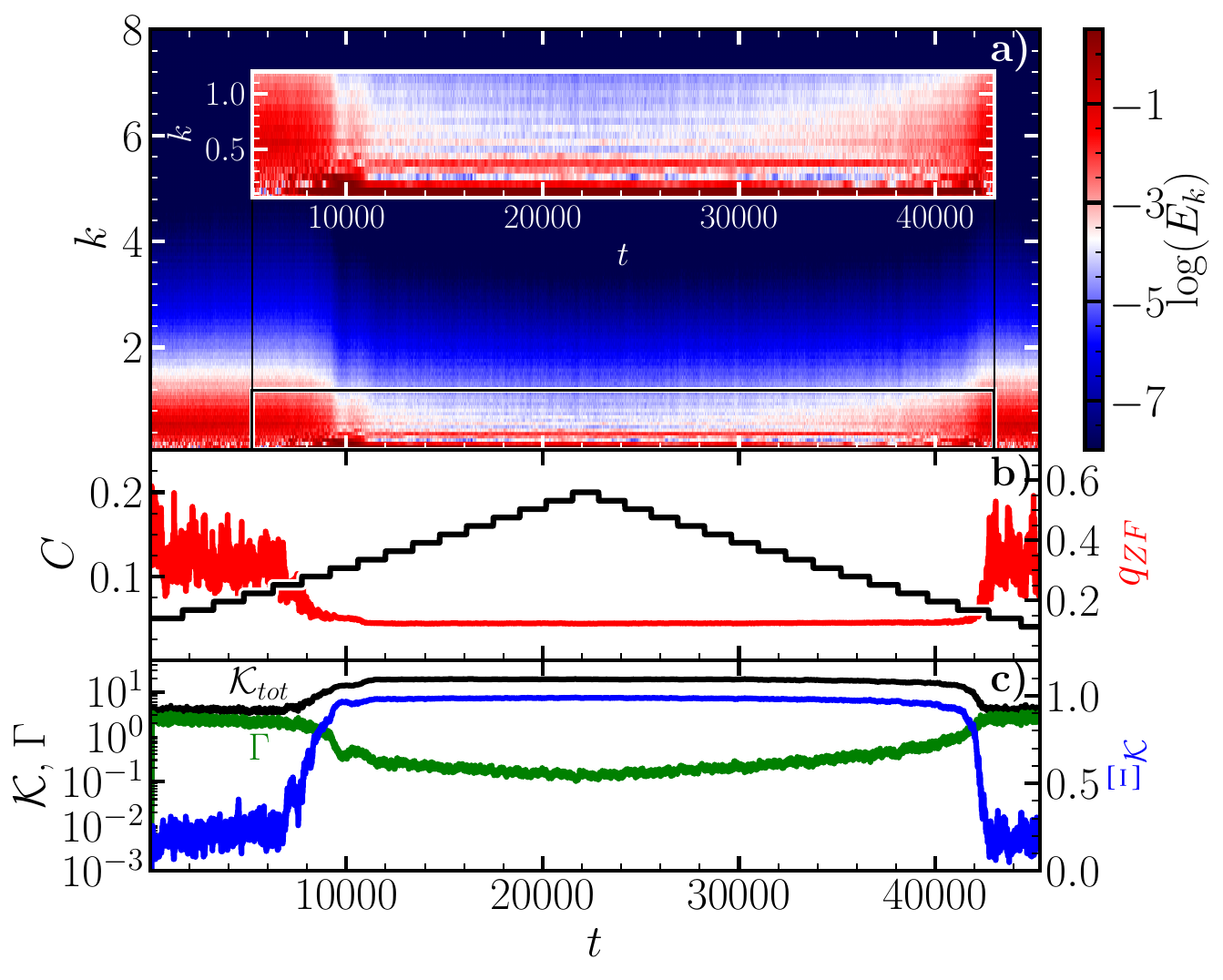}
\par\end{centering}
\caption{Several observables from the hysteresis simulation as functions of
time $t$. a) Logarithm of the kinetic energy spectrum $\log\left[E(k)\right]$
as a function of time $t$ ($x$-axis) and wave-number amplitude $k$
($y$-axis). b) adiabaticity parameter $C$ (black) and dominant zonal
mode $q_{ZF}$ (red). c) Total kinetic energy (black), radial particle
flux (green) and zonal energy fraction $\Xi_{K}$ (blue). \label{fig:Spectro-hyst}}
\end{figure}

\noindent In order to study the nature of the transition between 2D
turbulence and the quasi-1D zonal flow dominated state, we check if
there is a hysteresis behaviour when the adiabaticity parameter is
increased and then decreased around the transition point. For that
purpose, we launched a simulation at $C=0.05$, in the 2D turbulent
regime. After reaching saturation, we varied the adiabaticity parameter
by constant steps of $\Delta C=0.01$ until we reached $C=0.2$, well
above the threshold of zonal flow formation. Then we decreased the
adiabaticity parameter down to $C=0.04$, with the same increments
(see Figure \ref{fig:Spectro-hyst} middle plot for an illustration).
For each value of $C$, we let the system elvolve during $\Delta t(C)=200\times\gamma_{max}^{-1}(C)$
before changing the value of the adiabaticity parameter, which is
quite large compared to the energy injection time and allows the system
to respond to the modification of its linear properties. That way,
we perform an ``adiabatic'' transformation in the thermodynamic
sense, which allows the system to ``forget'' about its previous
state, in order to distinguish the hysteresis in the phase transition
from ``memory'' effects common in turbulence. Note also that the
size of the 2D box is kept constant at $L_{x}=L_{y}=32\pi$, while
the dissipation coefficients are varied with $C$ according to (\ref{eq:nu_dep}).

In Figure \ref{fig:Zonal-faction-hyst}, we show the zonal kinetic
energy (left) and enstrophy (right) fractions as functions of $C$
from this simulation. The blue triangles correspond to increasing
$C$ from 0.05 to 0.2, while the red triangles correspond to decreasing
$C$ from 0.2 to 0.04. Both fractions clearly evidence a hysteresis
loop, where two branches coexist, with two different critical values
of the adiabaticity parameter for the transition between 2D isotropic
turbulence and the zonostrophic regime. The lower branch, corresponding
to the transition from turbulence to zonal flows, has a critical value
around $C\approx0.09-0.1$. The upper branch, corresponding to the
transition from zonal flows to turbulence, has a lower critical value,
at $C\approx0.06$.

One can look a the hysteresis loop as a feature of a phase transition
between a hot disordered state (isotropic 2D turbulence) and a colder
organised state (zonal flows). In this framework, increasing $C/\kappa$
would amount to decreasing the available heat. On the other hand,
decreasing $C/\kappa$ in the quasi-1D state corresponds to increasing
the amount of energy of the system until the organised structures
collapse, like a transition from a solid crystal state to a liquid
state, which is endothermic. Because the collapse of the crystalline
structure needs to absorb some energy, the solid state, if formed,
can survive higher heat. But if not, one has to reduce heating in
order to form it. Applying such a picture could explain why the quasi-1D
state survives on the upper branch when we decrease $C/\kappa$ (\emph{i.e.}
increase the ``heating'') below the threshold of the lower branch,
suggesting that the collapse of the zonal flows requires some energy
absorption. There are also links to vortex crystal melting and percolation,
in the sense of turbulent eddies breaking through zonal flow barriers.

Moreover, an indirect indication of a possible time-scale divergence
around the transition point can be seen in the form of error bars
of the hysteresis loop. In Figure \ref{fig:Zonal-faction-hyst}, we
can see that the points inside the loop (\emph{e.g. }$C=0.09$, 0.10
of the lower branch, and $C=0.06$ of the upper branch) exhibit large
standard deviations. For these points, the system is initially in
the ``wrong branch'' and tries to go to the other one, in a time
that becomes very large when we are closer to the transition point.
A dedicated study to investigate the time-scale divergence of the
system towards the transition point could be performed, by starting
a simulation inside the hysteresis loop and measuring how long it
takes to jump from one branch to the other, as done in Refs. \onlinecite{kan:2019a,xu:2024},
but this is left for future work.

Note that contrary to the extensive adiabaticity parameter scan, where
each simulation is launched with a specific value of $C$, zonal flows
in the hysteresis run do not form through the modulational instability
that follows the initial linear growth of the dissipative drift-wave
instability. Here in contrast, zonal flows form in an already turbulent
system, in which the linear instability has somehow saturated. This
highlights the role of $C/\kappa$, which is a linear parameter, even
in the non-linear saturated state of the system, through various competing
mechanisms. It can be speculated that $C/\kappa$ plays this role
because the zonostrophy parameter $R_{\beta}$ is proportional to
it.

We can also observe the behaviour of several quantities such as the
energy spectrum, the dominant zonal mode, or the particle flux during
the hysteresis. In Figure \ref{fig:Spectro-hyst}, we show at the
top a) the time evolution of the logarithm of the ``isotropic''
kinetic energy spectrum $E(k)$. In the middle b), we plot the time
evolution of the adiabaticity parameter $C$ (black) and the dominant
zonal wave-number $q_{ZF}$ (red), computed using (\ref{eq:qZF}).
At the bottom c), we show the time evolution of the total kinetic
energy (black), the particle flux (green), and the zonal energy fraction
(blue). From these three plots, we can see that the transition from
turbulence to the zonal flow dominated state is associated with a
strong sharpening of the spectrum in the large scales, as evidenced
by the few emergent red lines (\emph{i.e.} dominant zonal modes),
while the other scales are dramatically suppressed. In contrast, as
$C$ is decreased, the zonal flows collapse and the spectrum rebroadens.
The horizontal red lines, which we see more clearly in the zoomed
panel, correspond to the zonal flows, whose energy spectrum is maximum
at $q_{ZF}\sim0.12-0.13$. The dominant zonal wave-number is almost
constant and independent of $C$ once the transition has occured,
as shown in plot b), even though the linear injection wave-number
$k_{y0}$ increases (see Figure \ref{fig:Maximum-linear-growth}).
Furthermore, this value is quite low compared to the typical zonal
wave-numbers obtained in the simulations where we fixed $C\sim0.1-0.2$,
as seen in Figure \ref{fig:qZF_C}. This is similar to the results
from Ref. \onlinecite{grander:2024} where the reformation of zonal flow
is studied after decreasing $C$ below the transition and re-increasing
it. In their case, the new zonal flows are larger than the original
ones which formed from a modulational instability. This is consistent
with the observation that, in this state, the zonal flows are formed
by a mechanism different from the modulational instability. It would
be interesting to continue increasing $C$ up to large values well
in the zonal flow dominated state ($C>1$), in order to study the
evolution of the zonal flows and observe their merging.

Finally, we discuss the impact of the transition on the total kinetic
energy and the particle flux, which are shown at the bottom c) of
Figure \ref{fig:Spectro-hyst}, respectively in black and green. As
expected, the formation of zonal flows reduces the radial particle
flux. On the other hand, the total kinetic energy is increased by
roughly an order of magnitude when zonal flows form, and is then decreased
when the flows collapse. We think that, since no dissipation or large-scale
friction is applied on zonal flows, they can store more energy than
the turbulent modes, which explains why the kinetic energy of the
system tends to be higher when they dominate.

\section{Transition in a reduced model \label{sec:Reductions}\label{sec:Reductions}}

In this Section, we investigate briefly how we can propose strong reductions of the
Hasegawa-Wakatani equations that can still capture
the transition from 2D turbulence to quasi-1D zonal flow dominated
state. The goal of such a study is actually two-fold: (i) it allows
one to determine which are the minimal physical features required
to see such a transition, and hence keep what is missing in models
that are too strongly recuded ; (ii) if a simple model is able to
reproduce the transition - or part of its key features - it can help
to provide a deeper understanding of the phenomenon, particularly
in finding theoretical explanations of its various aspects such as
the exact value of the transition point $C\approx0.1$, its dependence
on other parameters, such as viscosity, or the role played by the
adiabaticity parameter in the non-linear regime.

\begin{figure}[htbp]
\begin{centering}
\includegraphics[width=1\columnwidth]{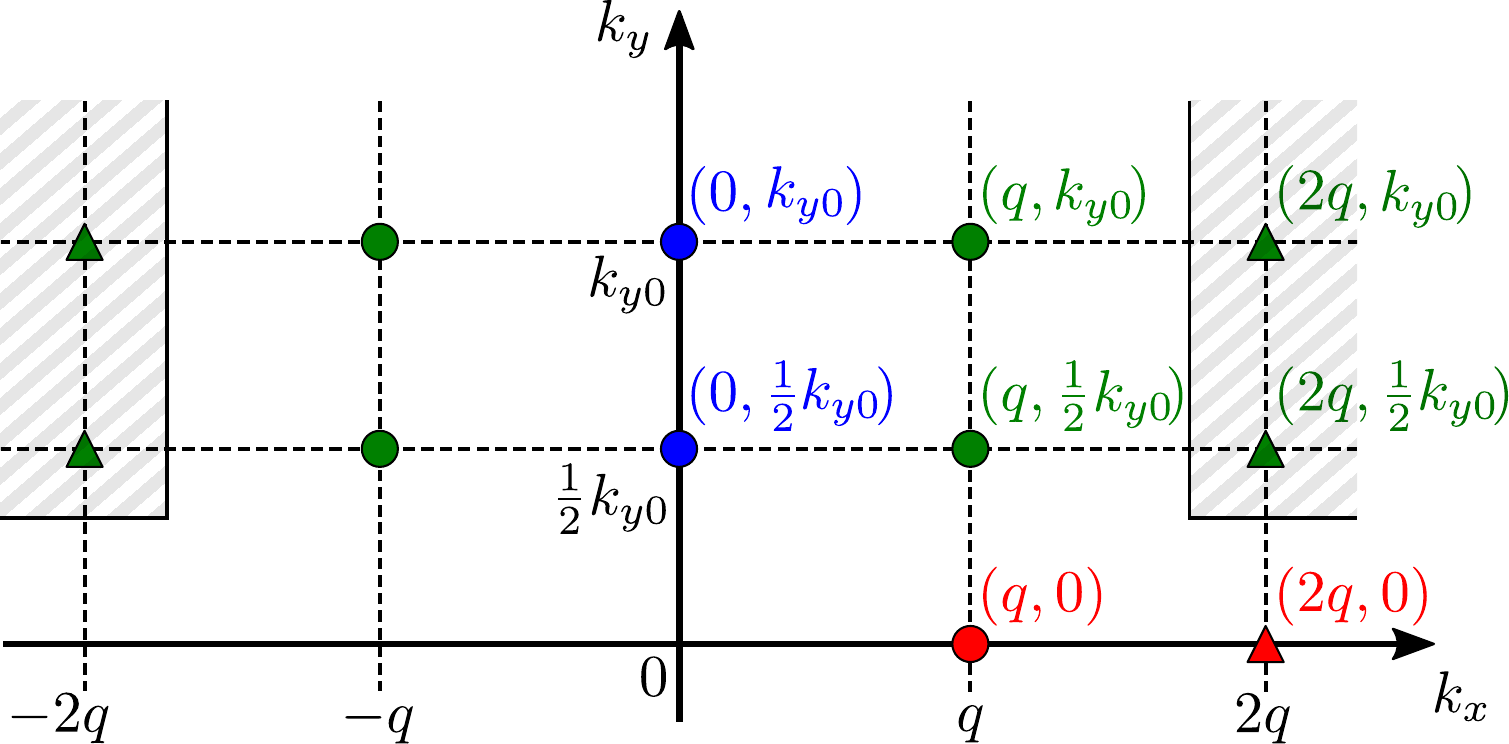}
\par\end{centering}
\caption{Schematic of the 8 modes (plus 4 buffer modes) reduction. The modes
in the $k_{x}<0$ region are symmetric to those in the $k_{x}>0$ region.
The dissipation acts only on the buffer modes, which are indicated
by the dashed rectangles. \label{fig:12-modes}}
\end{figure}

One such reduction can be performed using low order wave-number space
network models, which are the kind of low dimensional models of spectral
energy transfer \citep{terry:1983,gurcan:23}, and which have been
studied in detail for the Hasegawa-Wakatani system in Ref. \onlinecite{gurcan:2022}.
In the following, we detail the minimal wave-number space reduction
we found to be able to reproduce the transition. This reduction involves
retaining $2$ radial modes, \emph{i.e.} $(q,0)$ and ($2q,0$), and
$2$ poloidal modes, \emph{i.e.} $(0,k_{y0})$ and $(0,k_{y0}/2)$,
along with their corresponding inner {[}\emph{i.e.} $(\pm q,\frac{1}{2}k_{y0})$
and $(\pm q,k_{y0})${]} and outer {[}\emph{i.e.} $(\pm2q,\frac{1}{2}k_{y0})$
and $(\pm2q,k_{y0})${]} \emph{side-bands}, as shown in Figure \ref{fig:12-modes}.

The outer side-bands $(\pm2q,\frac{1}{2}k_{y0})$, $(\pm2q,k_{y0})$
(green triangles in dashed rectangles) are then strongly damped by
introducing rather large dissipation coefficients (\emph{i.e.} $\nu k^{2}\sim1$),
while the other modes are considered to be perfectly inviscid. In
other words these $4$ modes act as a ``\emph{buffer}'' zone to
dissipate the energy and the enstrophy, which is nonlinearly transferred
to those scales. Note that these are the modes that correspond to
a wave-number space grid used in the pseudo-spectral code with a $10\times10$
padded resolution. In reality, there are also triads with the modes
in the $k_{y}<0$ region and with zonal modes with $k_{x}<0$, but
these modes are accounted for using Hermitian symmetry: since $\phi(x,y,t)$
is a real quantity, we have $\phi_{-\mathbf{k}}=\phi_{\mathbf{k}}^{*}$.

\begin{figure*}[htbp]
\begin{centering}
\includegraphics[width=2\columnwidth]{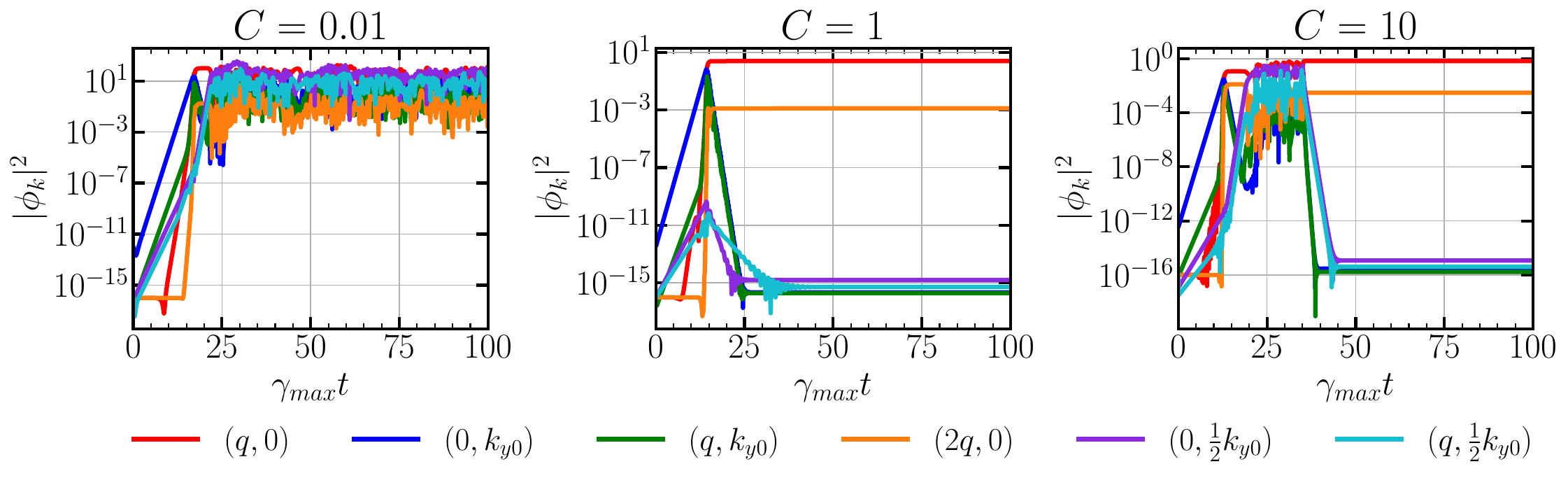}
\par\end{centering}
\caption{Time evolution of squared electrostatic potential amplitudes $|\phi_{k}|^{2}$
of 6 key modes in the 12 mode model, for $C=0.01$ (left), $C=1$
(center) and $C=10$ (right). The most unstable is in blue and the
zonal modes $(q,0)$ and $(2q,0)$ are respectively in red and orange.
The time is normalised by $\gamma_{max}$ (we show only the evolution
up to $t=100\times\gamma_{max}^{-1})$. \label{fig:DYN_12MODES}}
\end{figure*}

We solve the Hasegawa-Wakatani equations projected onto this reduced
Fourier space using the pseudo-spectral code, with different values
of $C$. We choose the zonal wave-number to be $q=0.6k_{y0}$, which
roughly corresponds to the zonal mode with the maximum growth rate
in the modulational instability framework applied to a single triad
in the limit $C\rightarrow +\infty$. Each simulation is run until $t=200\times\gamma_{max}^{-1}$.
The details on the rest of the parameters are given in Appendix \ref{sec:app-red-params}.
The time evolution of the amplitude $|\phi_{k}|^{2}$ of 6 Fourier
modes is shown in Figure \ref{fig:DYN_12MODES} for $C=0.01$, $C=1$
and $C=10$. The reduced model seems to behave similarly to the results
from Figure \ref{fig:energy_zonal_fraction}. For $C=0.01$ (left), turbulent
modes are not suppressed and coexist with the zonal modes (red and
orange), which suggests that the system is close to 2D isotropic turbulence.
On the contrary, for $C=1$ (center) and $C=10$ (right), the zonal
modes quickly dominate after their non-linear growths, which are followed
by a dramatic suppression of the turbulent modes (with a longer non-linear
interplay for $C=10$), while the amplitude of the zonal modes becomes
constant.

\begin{figure*}[htbp]
\begin{centering}
\includegraphics[width=1.5\columnwidth]{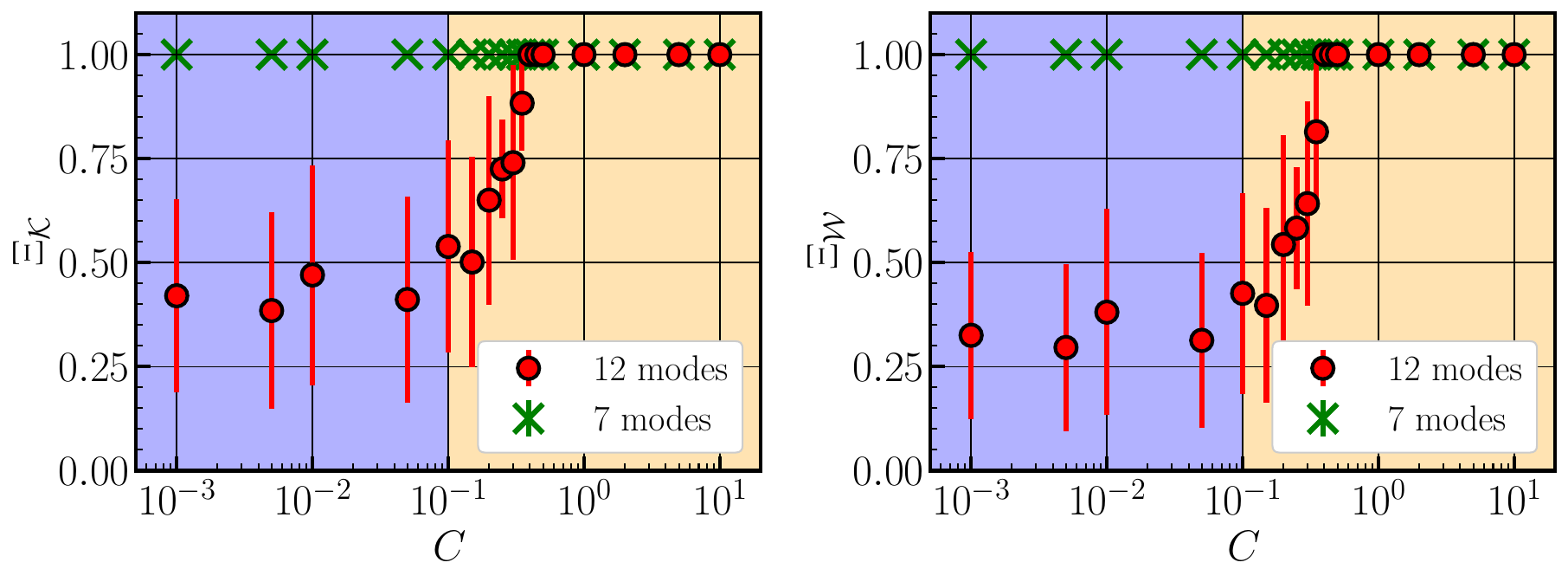}
\par\end{centering}
\caption{Zonal energy fraction (left) and enstrophy fraction (right) as functions
of $C$ for the 12 modes (red dots) and 7 modes (green crosses) reduced
models. Values are averaged over $t$ in the last quarter of the simulations.
Blue and yellow regions correspond respectively to $\Xi_{K}<50\%$
and $\Xi_{K}>50\%$ for the 12 modes reduction. \label{fig:ZF_C_reduced}}
\end{figure*}

We then investigate the eventuality of a transition from 2D isotropic
turbulence to zonal flows by varying the adiabaticity parameter over
the range $C\in[10^{-3},10]$, and we measure the zonal energy and
enstrophy fractions using (\ref{eq:zkf}) and (\ref{eq:zwf}). The
results are shown in Figure \ref{fig:ZF_C_reduced} (left: energy
fraction, right: enstrophy fraction, both averaged over the final
quarter of each simulation), where the red dots correspond to the
reduced model of 8 (plus 4) modes. Both fractions show the expected
transition, around $C\approx0.1$, even though we observe mainly a
separation between a regime where turbulent and zonal modes coexist
($C<0.4$), and another one where zonal modes dominate and turbulence
is completely suppressed ($C>0.4$). This observation may be linked
to the decaying of turbulence in the zonal flow dominated regime,
somewhat reminiscent of the Dimits shift \citep{dimits:00}, which
is illustrated on the center and right plots in Figure \ref{fig:DYN_12MODES},
that kills the turbulent modes when the zonal modes dominate. In contrast,
there is no such brutal suppression of turbulence close to the transition
in DNS, and the turbulent kinetic energy spectra is still high at
smaller scales, which is evidenced by smaller scale eddies surviving
and being advected by the large scale flows. This is not possible in our
reduced model, since it has only large scale modes.

We also performed this analysis with a 5 modes (plus 2 buffer modes)
model, which involves only the triadic interaction between the most
unstable mode, the zonal mode and the corresponding side-band\emph{s.}
In Figure \ref{fig:12-modes}, this corresponds to removing the mode
$(0,\frac{1}{2}k_{y0})$ and its associated side-bands. Such a model,
can allow ``simple'' analytical investigations, but as shown in
Figure \ref{fig:ZF_C_reduced} (green crosses), no transition has
been observed in this model, while varying the adiabaticity parameter
$C$. Looking at the dynamical evolutions of the mode amplitudes (not
shown here), all results were similar to the $C=1$ case from Figure
\ref{fig:DYN_12MODES}, with dominating zonal flows and decaying turbulence.
Nevertheless, for low values of $C$, we noticed a longer nonlinear
interaction between the zonal and turbulent modes, even though the
zonal mode ends up dominating and the turbulent modes decaying. We
suspect that in systems with a larger number of modes, which correspond
to higher resolution grids, some interactions involving multiple turbulent
modes may transfer part of the energy to the large scale turbulence,
instead of transferring it to the zonal flows, while at the same time
transfering part of the enstrophy to small scales, possibly through
a mechanism of the dual cascade. We believe that the adiabaticity
parameter $C$ might play a role in the competition between this mechanism
and the transfer of energy to zonal modes, maybe through some linear
process, or through the modification of the resonance manifold.

This interpretation is also consistent with another analysis we performed
with the 12 mode reduced model, using $(0,2k_{y0})$ instead $(0,\frac{1}{2}k_{y0})$
and all corresponding side-bands, (corresponding almost to the model
used in Ref. \onlinecite{terry:1983} except that we have two more additional
buffer modes). In this case, the results are similar to the 7 mode
reduction: zonal flows always dominate and we don't observe any transition.
This strongly suggests that one has to allow for the possibility of
an inverse cascade, in addition to the zonal flow generation mechanism,
in order to have two competing mechanisms necessary for the transition,
which is possible only if there is at least one smaller poloidal wave-number
in addition to $k_{y0}$. We also observed similar discrepancies
between 1D simulations (completely resolved in the radial direction)
with 1 mode in the $y$ direction (most unstable mode) since such
a model has the interactions with the zonal flows, but no possibility
of an inverse cascade in the classical sense.

It should also be noted that a preliminary
study attempting to reproduce the hysteresis loop with the reduced
model was inconclusive, or rather concluded that such an effort requires a
more careful approach, since varying $C$ in real time with only a few
modes, where one of them is the most unstable mode, results in either
having to change the box size in real time, or including a less reduced
model for the large scales.

Finally, we emphasise that the performance of the 12 mode reduced
model relies strongly on its different parameters, including the phases
of its \emph{seed} initial conditions (detailed in Appendix \ref{sec:app-red-params}).
Since it contains only a few modes, phase synchronisation can significantly
affect its behaviour, leading to variations of zonal fractions between
different runs for the same input parameters. In order to mitigate
that effect, we increased the initial amplitude of the most unstable mode
by a factor 100 (still much smaller than the saturation levels as
seen in Fig. \ref{fig:DYN_12MODES}), so that it is forced to act
as a ``pump'' of energy in the initial phase of the system and ensure
that the initial part of the nonlinear coupling between the modes
remains ``well-behaved''. The numerical value of the dissipation
imposed on the buffer modes can also affect how the model behaves,
which is probably expected since high levels of small scale dissipation
would kill the buffer modes and may force the system to interact with
the zonal modes. Lastly, in the high $C$ regime, we observe some
resurgence of the turbulent modes long after the ``turbulence decay''.
This could be due to phase synchronisation effects, possibly similar
to the plasma echo phenomenon, which would hopefully diminish with
the system size. All these observations require a deeper investigation
(along with some analytical studies) of the dependencies of the model
on its various parameters, and to understand whether or not the results
can be extended to the observations made in the DNS. Particularly,
the effect of dissipation on the transition in both DNS and reduced
model should be carefully studied.

\section{Conclusion}

In this article we studied, in some detail, the transition from 2D
isotropic turbulence to a quasi-1D state dominated by zonal flows
in the Hasegawa-Wakatani system, when the adiabaticity parameter $C$
(or $C/\kappa$ when the system is properly normalised) is varied.
In particular, we observed a transition point around $C\approx0.1$
using zonal kinetic energy and enstrophy fractions as order parameters.
While the zonal energy fraction exhibits a sharp transition from low
values (corresponding to isotropic turbulence) to almost $100 \%$ (zonal
flow dominated system), the zonal enstrophy fraction increases only
gradually at the transition point. This can be explained by the fact that, while the energy
is concentrated at large scales, which are dominated by zonal flows,
the enstrophy spectrum remains important for smaller scales, as witnessed
by small scale eddies living inside the zonal flows for $C\in[0.1,1]$.

We proposed the interpretation that this transition can be seen as
a change from a 2D strongly turbulent system where 2 flow quantities
(energy and enstrophy) are conserved by the non-linear terms, to a
weakly turbulent system, dominated by waves, for which there is a
third conserved quantity, the zonostrophy, in the asymptotic limit
($C\rightarrow+\infty)$ for the resonant interactions among waves.
It may be that, as in the case of the transition from forward to inverse
cascade in three dimensional turbulence, the conservation of the new
quantity in the asymptotic limit modifies the behaviour of the system
even before it is fully conserved.

In numerical simulations where the adiabaticity parameter has been
varied across the transition point, we observed a clear hysteresis
loop for both zonal kinetic and enstrophy fractions. We noted that
this observation can be linked to phase transitions in various other
systems, where the disordered ``hot'' state corresponds to 2D isotropic
turbulence and the ``colder'' organised state corresponds to the
zonal flow dominated one. The equivalent of ``heating'', or more
accurately ``cooling'', is shown to be linked to the control parameter
$C/\kappa$, which determines the energy injection through its effect
on the linear instability. The presence of the hysteresis suggests
that, similarly to ice melting or water boiling, zonal flows collapsing
requires some latent heat in order to break the organised structure
of this state, suggesting a first order phase transition. Moreover,
the fact that we observed the generation and then the collapse of
zonal flows in a turbulence saturated system highlights that $C/\kappa$
plays a role in favouring energy or enstrophy transfers either towards
small scale turbulence, where they are dissipated, or towards zonal
flows, where they remain, the detailed mechanism of which needs to
be identified.

We also studied the radial particle flux and found that it decreases
with increasing $C$ following an approximate power-law, whose exponent
changes between 2D isotropic turbulence ($\Gamma\propto C^{-0.35}$
for $C\in[10^{-4},1.3\times10^{-1}]$) and the zonal flow dominated
states ($\Gamma\propto C^{-2}$ for $C\in[1.3\times10^{-1},2]$),
though the flux departed slightly from the power law for $C>2$. It
was found that this trend can be reproduced, both qualitatively and
quantitatively using a simple saturation rule accounting for zonal
flows which can be written as $\Gamma=\Gamma_{sat}\left(1-\Xi_{K}\right)$.

Finally, we reproduced the transition using a minimal wave-number
space network model. The fact that further reduced models fail to
exhibit the transition highlights the importance of the triadic interactions
involving only turbulent modes of scales larger than the energy injection
scale (\emph{i.e.} poloidal modes with $k_{y0}/2$). This key aspect could
be inserted in other future reduced models of the Hasegawa-Wakatani
system (and other instability-driven turbulence models). However,
detailed studies on the dependency of our model on its various parameters
are required, especially on the role of dissipation on the transition,
which also seems to impact results from DNS. Nevertheless, the analytical
investigation of the 12 mode reduced model could provide a first
explanation of the role played by $C/\kappa$ in the competition between
zonal flows and 2D isotropic turbulence. One path to follow would
be to understand the so-called ``turbulence decay'' that we observed
when zonal flows become constant in the reduced model, which can be
thought of as a partial fixed point of the system.

Another following task would be to check whether self-consistent quasi-linear
and generalised quasi-linear reductions \citep{marston:16,nivarti:2024}
of the Hasegawa-Wakatani equations are able to reproduce the transition.
While we expect the self-consistent quasi-linear reduction to fail,
where fluctuation-fluctuation interactions are completely neglected,
one would guess that a generalised quasi-linear reduction, which includes
fully nonlinear evolution of the scales up to the injection scale,
should be able to reproduce the transition.
\begin{acknowledgments}
This work was granted access to the Jean Zay super-computer of IDRIS
under the allocation AD010514291 by GENCI. The authors would like
to thank the Isaac Newton Institute for Mathematical Sciences, Cambridge,
for support and hospitality during the programme “Anti-diffusive dynamics:
from sub-cellular to astrophysical scales”, and particularly mention
the workshop ``Mathematical and computational modelling of anti-diffusive
phenomena'' from which was inspired this work. This work has been carried out within the framework of the EUROfusion Consortium, funded by the European Union via the Euratom Research and Training Programme (Grant Agreement No 101052200 — EUROfusion) and within the framework of the French Research Federation for Fusion Studies. 
\end{acknowledgments}

\appendix

\section{Eigenvalues of the Hasegawa-Wakatani system \label{sec:app-eigenvalues}}

Linearisation and Fourier transform of the Hasegawa-Wakatani equations
for non-zonal modes $(k_{y}\neq0)$ yields
\begin{align*}
\partial_{t}\phi_{k}+(A_{k}-B_{k})\phi_{k} & =\frac{C}{k^{2}}n_{k}\\
\partial_{t}n_{k}+(A_{k}+B_{k})n_{k} & =(C-i\kappa k_{y})\phi_{k}\text{ ,}
\end{align*}
 where
\begin{align*}
A_{k}= & \frac{1}{2}\left[(Dk^{2}+C)+\left(\frac{C}{k^{2}}+\nu k^{2}\right)\right]\,,\\
B_{k}= & \frac{1}{2}\left[(Dk^{2}+C)-\left(\frac{C}{k^{2}}+\nu k^{2}\right)\right]\,.
\end{align*}

The two eigenvalues $\omega_{k}^{\pm}(C,\kappa,\nu,D)=\omega_{k,r}^{\pm}+i\gamma_{k}^{\pm}$
can then be written as
\[
\omega_{k}^{\pm}=\Omega_{k}^{\pm}-iA_{k}\,
\]
 with
\[
\Omega_{k}^{\pm}=\pm\left(\sigma_{k}\sqrt{\frac{H_{k}-G_{k}}{2}}+i\sqrt{\frac{H_{k}+G_{k}}{2}}\right)\,,
\]
 where $\sigma_{k}=\text{sign}(\kappa k_{y})$,
\[
H_{k}=\sqrt{G_{k}^{2}+C^{2}\kappa{{}^2}k_{y}^{2}/k^{4}}\,,
\]
and
\[
G_{k}=\left(B_{k}+\frac{C^{2}}{k^{2}}\right)\,.
\]

\section{Parameters of the 12 mode reduced model \label{sec:app-red-params}}

We choose the zonal wave-number to be $q=0.6k_{y0}$, which roughly
corresponds to the mode with the maximum growth rate in the modulational
instability framework applied to a single triad in the limit $C\rightarrow +\infty$,
but other values of $q<k_{y0}$ seem to work too (\emph{e.g.} $q=k_{y0}/2$).

The dissipation term, which is only applied on the buffer modes, is
$\nu k^{2}=10\times(\gamma_{max}\omega_{r,max})^{\frac{1}{2}}$, where
$\omega_{r,max}$ is the real frequency of the eigenvalue associated
to the maximum growth rate $\gamma_{max}$. The motivation to use
such ``hybrid'' inverse time is the observation that, while for
lower values of $C$ the dissipation term $\nu k^{2}$ roughly compensates
the injection $\gamma_{max}$ (with some empirical numerical prefactor),
for large values of $C$ we have $\omega_{r,max}\gg\gamma_{max}$,
which means that the linear process dominating in this regime is the
wave propagation and not the linear growth (though one can use some
other form that saturates for small growth rates).

Finally, all initial amplitudes are set to $10^{-8}$ and all initial
phases are random, with the exception of the most unstable mode $(0,k_{y}^{max})$
for which the initial amplitude is set to $10^{-6}$ (still well below
its saturation level). This guarantees that the most unstable mode
plays the role of a ``pump'', injecting energy in the system. This
causes the system to have a ``well-behaved'' linear phase and suppresses
the effect of unwanted phase synchronisations, which tends to appear
randomly depending on the phases of the seed initial conditions. Each
simulation is run until $t=200\times\gamma_{max}^{-1}$. The details
of the parameters used for the simulations are summed up in Table
\ref{tab:triad_params}.

\begin{table}[H]
\centering{}%
\begin{ruledtabular}
\begin{tabular}{cccc}
\addlinespace[0.1cm]
$q$ & $\nu k^{2}$ (only on buffer) & $|\phi_{0,k_{y0}}|_{{\large |}{\large t=0}}$ & $|\phi_{k_{x},k_{y}\neq k_{y0}}|_{{\large |}{\large t=0}}$\tabularnewline\addlinespace[0.1cm]
\midrule 
\addlinespace[0.1cm]
0.6$k_{y0}$ & $10\times\left(\gamma_{max}\omega_{r,max}\right){}^{\frac{1}{2}}$ & $10^{-6}$ & $10^{-8}$\tabularnewline\addlinespace[0.1cm]
\end{tabular}
\end{ruledtabular}
\caption{Parameters used for the reduced model: zonal mode $q$, dissipation
$\nu k^2		$ applied on buffer modes and initial amplitude. Note that $\omega_{r,max}$
is the real part of the eigenvalue with maximum growth rate.\label{tab:triad_params}}
\end{table}

%

\end{document}